# On-demand Photonic Ising Machine with Simplified Hamiltonian Calculation by Phase Encoding and Intensity Detection

Jiayi Ouyang[1], Yuxuan Liao[1], Zhiyao Ma[1], Deyang Kong[1], Xue Feng[1]*, Xiang Zhang[2], Xiaowen Dong[2], Kaiyu Cui[1], Fang Liu[1], Wei Zhang[1], and Yidong Huang[1]

[1]Department of Electronic Engineering, Tsinghua University, Beijing 100084, China
[2]Central Research Institute, Huawei Technologies Co., Ltd., Shenzhen, 518129, China
*Correspondence: x-feng@tsinghua.edu.cn

**Abstract:** The photonic Ising machine is a new paradigm of optical computing that takes advantage of the unique properties of light wave propagation, parallel processing, and low-loss transmission. Thus, the process of solving combinatorial optimization problems can be accelerated through photonic/optoelectronic devices, but implementing photonic Ising machines that can solve arbitrary large-scale Ising problems with fast speed remains challenging. In this work, we have proposed and demonstrated the Phase Encoding and Intensity Detection Ising Annealer (PEIDIA) capable of solving arbitrary Ising problems on demand. The PEIDIA employs the heuristic algorithm and requires only one step of optical linear transformation with simplified Hamiltonian calculation by encoding the Ising spins on the phase term of the optical field and performing intensity detection during the solving process. As a proof of principle, several 20 and 30-spin Ising problems have been solved with high ground state probability (≥0.97/0.85 for the 20/30-spin Ising model).

## Introduction

Optimization problems[1] are ubiquitous in nature and human society, such as ferromagnets[2,3], phase transition[4], artificial intelligence[5], finance[6], biology[7,8], agriculture[9], *etc.*. Usually, combinatorial optimization problems (COPs) are non-deterministic polynomial hard (NP-hard) problems[1], in which the required resources to find the optimal solutions grow exponentially with the problem scales on conventional von-Neumann machines. To tackle such obstacles, Ising machines as specialized hardware-sovlers are introduced to accelerate the solving process[10], since any problem in the complexity class NP can be mapped to an Ising problem within polynomial complexity[11–13]. The NP-hard Ising problems can be described as finding the energy ground state corresponding to a specific Ising spin vector $\boldsymbol{\sigma}$, which is equivalent to searching the global minima of Hamiltonian in absence of the external field[14]

$$H(\boldsymbol{\sigma}) = -\frac{1}{2}\sum_{1\leq i,j \leq N} J_{ij}\,\sigma_i\sigma_j = -\frac{1}{2}\boldsymbol{\sigma}^{\mathrm{T}}\mathbf{J}\boldsymbol{\sigma}, \tag{1}$$

where $\mathbf{J}$ is the adjacent interaction matrix, $\boldsymbol{\sigma} = [\sigma_1, \sigma_2, \ldots, \sigma_N]^{\mathrm{T}}$ is the Ising spin vector including Ising spins $\sigma_i \in \{1, -1\}$, and the superscript T denotes the transpose. To implement Ising machines, various classical and quantum physical systems have been employed. For the quantum annealers, various systems, such as D-wave systems[15], trapped atoms[16,17], magnetic tunnel junctions[18] and radio-frequency superconducting quantum interference devices[19], have been employed to simulate the Ising model. However, the large-scale implementation and fault tolerance of the quantum system are still technically challenging, and the annealing process should be sufficiently slow to maximize the ground state probability[10]. In classical domain, the Ising machines mainly based on electronic devices can be divided into two categories according to the operation mechanisms. In the first category, the Ising spins are represented by the phase of coupled electronic oscillators and the system is driven to the ground state[20]. In the other category, electronic parallel computation schemes, such as the memristor crossbar[21] or the complementary metal-oxide-semiconductor chip[22,23], are utilized to accelerate the Hamiltonian calculations. Recently, advanced photonics exhibit the feasibilities to encode, transmit and process information on various spatial degrees of freedom, *e.g.* phase, amplitude/intensity, frequency/wavelength, time slot and spatial profile/distribution. Thus, photonic Ising machines have emerged due to the nature of parallelism and high propagation speed of light. Likewise, there are also two types of photonic Ising machines that have been experimentally reported. The first type of photonic Ising machine is based on the nonlinear optical or optoelectronic parametric oscillators[24–29], in which Ising spins are encoded on the phase terms of the time-multiplexed pulses and the ground state search relies on the spontaneous convergence of the parametric oscillators. For these Ising machines, the coupling among Ising spins has to be implemented via measurement feedback and performing vector-matrix multiplication in an electronic hardware, or aligning the oscillator pulses in the same time slot with the optical delay line. Moreover, whether Ising machines based on nonlinear parametric oscillators can solve random Ising models remains contested[30]. The second type of photonic Ising machine employs heuristic algorithms to solve Ising problems, in which the main computation task is to calculate the Ising Hamiltonians of different Ising spin states iteratively. Such task can be regarded as multiplying a static matrix by an ever-changing vector, which is very suitable for the optical vector-matrix transformation (OVMM) system. Actually, the OVMM has been applied on some specific computation tasks, such as the deep diffractive neural network[31,32] and the nanophotonic processor[33]. As shown in these works, the vector-matrix multiplications could be dramatically accelerated by photonic systems due to the parallel propagation of light. In the spatial photonic Ising machine[34], numerous Ising spins are encoded on the phase terms of the light field through phase-modulation units of the spatial light modulator (SLM). However, the spins interact in the intensity distribution of the light field with a Fourier lens, hence only some specific Ising models can be solved. The on-chip array of tunable Mach-Zehnder interferometers (MZIs) network[35,36], which is known as the Reck scheme[37], is also employed to solve arbitrary Ising models. However, due to the complexity of the Reck scheme, only 4-spin Ising models are experimentally demonstrated[36]. Therefore, it is still highly desired to implement photonic Ising machines that can solve arbitrary large-scale Ising problems with fast speed.

In this work, a photonic Ising machine is proposed and demonstrated with a fully reconfigurable OVMM system and the employed heuristic algorithm is modified from simulated annealing[38,39]. Meanwhile, the calculation of the Hamiltonian is simplified in the optical domain where the Ising spins are encoded on the phase term of the light field and only intensity detection is required. Our proposal is named as the "Phase Encoding and Intensity Detection Ising Annealer" (PEIDIA), which will be briefly explained as follows. At the beginning, with proper treatment of the adjacent matrix **J**, the calculation of Ising Hamiltonians can be modified from the quadratic form (Eq. (1)) and only one OVMM is required. Then with a simple summation of the intensities of the light field, the Hamiltonian can be readily calculated. At last, the heuristic algorithm is employed to search the ground state with the obtained Hamiltonian. Thus our proposed PEIDIA is quite helpful to simplify and speed up the calculations in the optical domain. Furthermore, the PEIDIA can serve as a kind of "on-demand" solver for arbitrary Ising models while a programmable OVMM setup is employed to perform arbitrary linear transformation of the input Ising spin vector.

In our experimental implementation, the employed OVMM scheme is based on the discrete coherent spatial (DCS) mode and SLMs, which is improved from our previous work[40–43]. To verify the feasibility of our proposal, we have experimentally solved an antiferromagnetic Möbius-Ladder model as well as two fully connected, randomly generated spin-glass models. For the 20-spin Möbius-Ladder model, the ground state probability reaches 0.99 (100 runs) within 400 iterations, while that of the fully connected and random model is around 0.97 (100 runs) within 600 iterations. Furthermore, the ground state probability of the 30-spin fully connected and random model is around 0.85 (100 runs) within 1200 iterations. It should be mentioned that, this proposed architecture does not rely on specific optical linear transformation schemes and heuristic algorithms. The main advantage of the PEIDIA is the simplified Hamiltonian calculation with only one OVMM so that the parallelism and fast-speed of the optical calculation may be fully exploited. Thus, the proposed PEIDIA would pave the way to achieve large-scale photonic Ising machines that can solve arbitrary Ising models on demand.

**Results**

**Architecture and operation principles of the PEIDIA.** In order to accelerate the solving process of the Ising problem, we have proposed an on-demand photonic Ising machine and Figure 1a shows the architecture design. There are three main stages in the operation of the PEIDIA: the electronic pretreatment, the optical matrix multiplication and the electronic feedback. In the first stage of the electronic pretreatment, the parameters for the setup of the optical system are calculated and configured according to the adjacent matrix **J** of the given Ising model. In the second stage, the optical matrix multiplication system accelerates the calculation of the Hamiltonian of a certain Ising spin vector. The optical intensities obtained from the optical matrix multiplication are detected and converted to electronic signals. In the third stage, the spin vector for the next iteration will be generated and fed to the optical matrix multiplication according to the adopted heuristic algorithm. The second and the third stage will be conducted iteratively until the algorithm terminates. The details are described as follows.

The main purpose of the electronic pretreatment is to simplify the calculation in the optical domain. According to Eq. (1), the Ising Hamiltonian has a quadratic form and two steps of vector-matrix multiplications are required. Actually, only one vector-matrix multiplication is needed in the optical domain with proper pretreatment. First, every real interaction matrix **J** of the given Ising model can be decomposed to a symmetric matrix $\mathbf{J}_+ = (\mathbf{J} + \mathbf{J}^T)/2$ and an anti-symmetric matrix $\mathbf{J}_- = (\mathbf{J} - \mathbf{J}^T)/2$. Since $\mathbf{J}_-$ has no contribution to the Ising Hamiltonian in Eq. (1), we only discuss the symmetric component $\mathbf{J}_+$ and use **J** to denote $\mathbf{J}_+$ for simplicity in this article. As **J** is a real symmetric matrix, the Hamiltonian has the form as follows with eigen-decomposition[44]:

$$H(\boldsymbol{\sigma}) = -\frac{1}{2}\boldsymbol{\sigma}^T \mathbf{J} \boldsymbol{\sigma} = -\frac{1}{2}\boldsymbol{\sigma}^T (\mathbf{Q}^T \sqrt{\mathbf{D}} \sqrt{\mathbf{D}} \mathbf{Q}) \boldsymbol{\sigma} = -\frac{1}{2}(\mathbf{A}\boldsymbol{\sigma})^T(\mathbf{A}\boldsymbol{\sigma}) \qquad (2)$$

where $\mathbf{J} = \mathbf{Q}^T \mathbf{D} \mathbf{Q}$, while **Q** is the normalized orthogonal eigenvector matrix and $\mathbf{D} = \text{diag}(\lambda_1, \lambda_2, \ldots, \lambda_N)$ is the diagonal eigenvalue matrix of **J**.

In our proposal, the vector-matrix multiplication of $\mathbf{A}\boldsymbol{\sigma}$ is performed by the OVMM. During the ground state search, the transformation matrix **A** is unchanged while the sampled spin state **σ** updates iteratively. Thus, each Ising spin is considered as encoded on the phase term of the optical field $E_i = E_0 \sigma_i = E_0 \exp(i(\varphi_0 + \varphi_i))$, while $\varphi_i$ corresponds to the element of the spin vector with the value of $\varphi_i \in \{0, \pi\}$. By neglecting the constant phase term of $\exp(i\varphi_0)$, the complex amplitude of the output optical field can be written as

$$\mathbf{E} = E_0 \mathbf{A} \boldsymbol{\sigma}, \tag{3}$$

where $E_0$ is the constant amplitude term. By defining the output intensity vector **I** by $I_i = E_i^* E_i$, the Hamiltonian becomes (see detailed deduction in Supplementary Note 1)

$$H = -\frac{\mathbf{E}^T \mathbf{E}}{2E_0^2} = \frac{1}{2E_0^2}\left(\sum_{i,\lambda_i<0} I_i - \sum_{i,\lambda_i>0} I_i\right). \tag{4}$$

Eq. (4) shows that the calculation of Hamiltonian turns into the simple summation of the optical intensities in a subtle way. Thus in our proposal, the optical computation would perform the task of encoding spin vectors on optical field, vector-matrix multiplications of **Aσ** and intensity detections as shown in Figure 1a. It should be mentioned that although the Ising spin vector **σ** is encoded on the phase term of optical field, only the measurement of the output intensity vector **I** is required to obtain the Hamiltonian.

In Eq. (4), the first term in the bracket is the summation of the intensities corresponding to the negative eigenvalues, while the second term is that corresponds to the positive eigenvalues, which is due to the difference between $E_i E_i$ and $E_i^* E_i$ when $\lambda_i < 0$. Since there are subtracting operations, the Hamiltonian is finally calculated with Eq. (4) in the electronic domain after the intensity detection. In succession, the heuristic algorithm determines the spin vector for next iteration. Here, a modified simulated annealing algorithm is employed to search the ground state. In the iteration $n$, a spin state $\boldsymbol{\sigma}^{(n)}$ is accepted and its Hamiltonian $H^{(n)}$ is calculated. Then in the next iteration $(n + 1)$, $m$ spins of the spin vector are randomly flipped, which means the spin vector $\boldsymbol{\sigma}^{(n)}$ is updated to $\boldsymbol{\sigma}^{(n+1)}$ on the optical domain via the electronic feedback. The variable $m$ is a random integer obtained from a Cauchy random variable $C(0, \alpha T)$, where $\alpha$ is a scaling coefficient and $T$ is the annealing temperature (see Supplementary Note 2). With such state-generation method, the spin state can experience long jumps occasionally and escape from local minima more easily. Then the difference between the current Hamiltonian $H^{(n+1)}$ and the previous one of $H^{(n)}$ is calculated as:

$$\Delta H = H^{(n+1)} - H^{(n)}. \tag{5}$$

If $\Delta H \leq 0$, the generated state $\boldsymbol{\sigma}^{(n+1)}$ is accepted. If $\Delta H > 0$, $\boldsymbol{\sigma}^{(n+1)}$ is accepted with the probability of $\exp(-\Delta H/T)$ due to the Metropolis criterion[38,39]. In a single run of the algorithm, $T$ is slowly decreased from the initial temperature of $T_0$ to zero with the increase of the iteration number. According to the annealing schedule, it can be seen that in the early stage of the annealing, the PEIDIA can perform the global search, while at the end of the annealing, it is more likely to perform the local search. Finally, a "frozen" state will be obtained, which may be the optimal ground state with high probability. Actually, other heuristic algorithms can also be adopted, such as genetic algorithm[45], in which multiple-spin-flips are also desired.

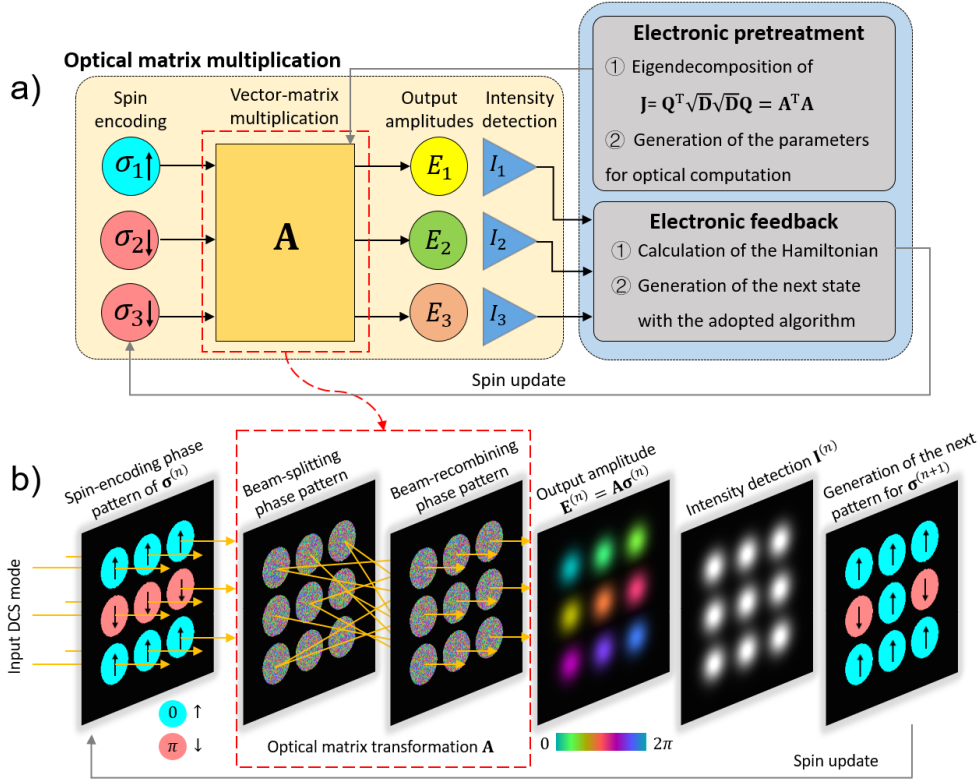

**Figure 1. The architecture diagram of the Phase Encoding and Intensity Detection Ising Annealer (PEIDIA). a)** The full operation process of the PEIDIA consisting of three main stages. $\sigma_i$: the Ising spins encoded on the phase term of the light field. **A**: the transformation matrix. $E_i$: the output complex amplitude. $I_i$: the output intensity. **J**: the interaction matrix of the Ising model. **b)** Detailed demonstration of the PEIDIA with arbitrary non-unitary matrix transformation of the discrete coherent spatial mode for a 9-spin Ising problem. The input field consists of 9 Gaussian beams with equal intensities. Different colors in wavefront modulation patterns and optical fields indicate different phases according to the color bar, while the black region denotes 0 phase delay for readability. In the insets of the optical fields, the brightness denotes the optical amplitude. The orange arrows denote the propagation directions of the Gaussian beams. In the shown spin-update process, one red circular region turns blue, representing a spin flip.

**Optical vector-matrix multiplication.** Generally, the transformation matrix **A** in Eq. (3) is complex and non-unitary, so that the OVMM employed in our architecture should be capable of achieving such non-unitary transformation. In our previous work[40–43], a matrix transformation scheme has been demonstrated with discrete coherent spatial (DCS) mode and SLMs. Such scheme can perform arbitrary complex vector-matrix multiplications for both unitary and non-unitary matrices. Based on it, the architecture of the optical computation in the PEIDIA is schematically depicted in Figure 1b. The spin vector is encoded on the input DCS mode, which consists of a group of Gaussian beams. More specifically, the elements of the spin vector are defined as the complex amplitudes at the centers of the Gaussian beams respectively. During the annealing process, each Ising spin $\sigma_i^{(n)}$ is encoded on the input vector through the appending spin-encoding phase pattern corresponding to the phase delay of $0/\pi$, as depicted by the blue/red circular regions in Figure 1b, respectively. Then the input vector passes through the meticulously designed beam-splitting and recombining phase patterns which are determined by the transformation matrix **A** (see details in Supplementary Note 3). Once generated for a given Ising problem, such two patterns are fixed during the following calculation and annealing process. The output amplitude vector $\mathbf{E}^{(n)}$ consists of the complex amplitudes at the centers of the beams in the output plane, where the output intensity vector $\mathbf{I}^{(n)}$ is detected. In succession, the Hamiltonian $H^{(n)}$ is calculated in the electronic domain and the next sampling state $\boldsymbol{\sigma}^{(n+1)}$ is generated and updated to the spin-encoding phase pattern. The spin flip can be simply achieved by adding a constant phase delay $\pi$ to the corresponding circular region of the spin-encoding phase pattern, as depicted in the last pattern of Figure 1b.

It should be mentioned that with our scheme, the mapping relations between the Ising model and the experimental parameters (the phase patterns corresponding to the matrix **A**) are simple and explicit. The beam-splitting and recombining scheme can directly conduct the non-unitary matrix transformation without utilizing cascaded structures[40,41].

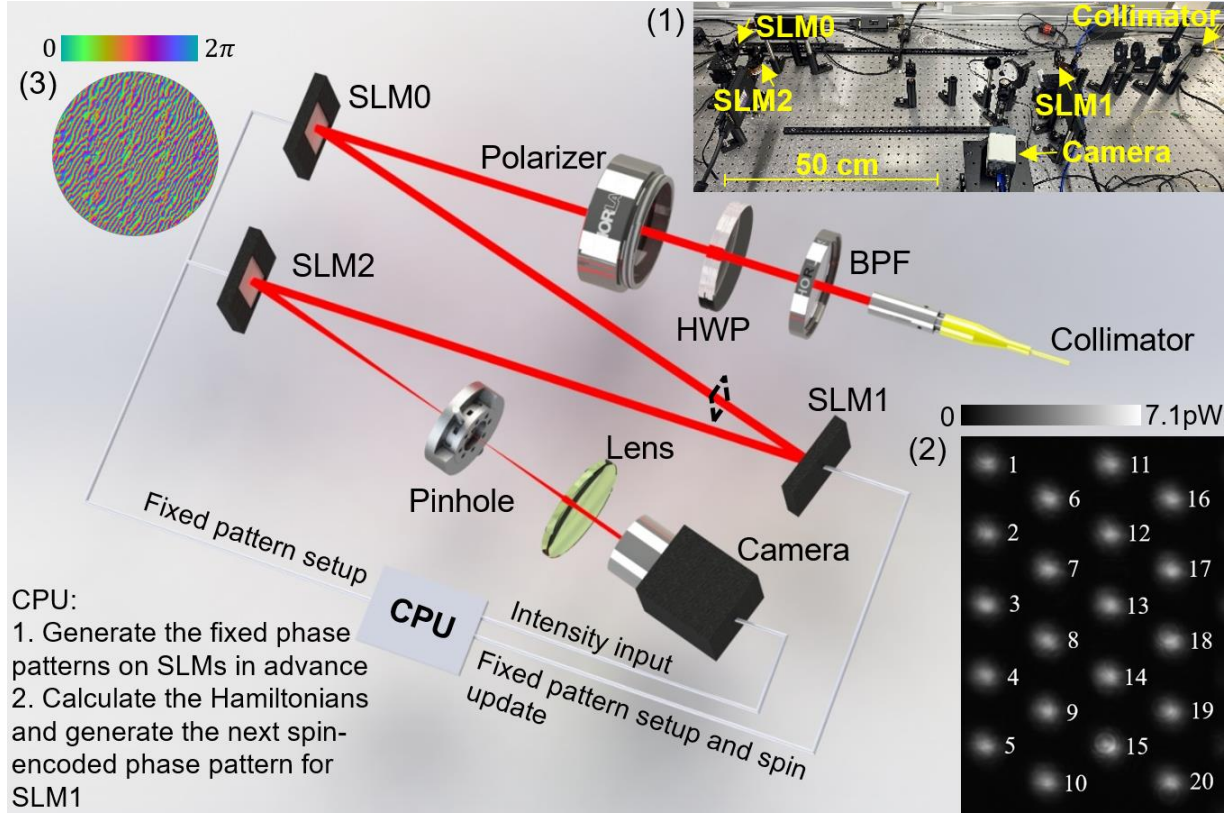

**Figure 2. The experimental setup.** BPF: bandpass filter. HWP: half-wave plate. As the employed spatial light modulators (SLMs) are reflective, additional blazed gratings are applied to all SLMs to extinct the zero-order diffractions. Inset (1) is the photograph of the experimental demonstration and the detailed description is provided in Supplementary Note 3. Inset (2) shows the intensity distribution in the transverse plane surrounded by the black dashed box. Here SLM0 splits the beam into $N$=20 beams, which is equal to the number of the Ising spins. The color bar denotes the incident power on each pixel. Inset (3) shows a circular beam-recombining phase pattern on SLM2 which is not superposed with the blazed grating, and the color bar shows the phase distribution. SLM1 encodes the Ising spins to the beams, and implements the optical vector-matrix multiplication together with SLM2.

**Experiment demonstration.** The experimental setup of the PEIDIA with 20 spins is illustrated in Figure 2, and the photo of the experimental setup is shown in inset (1) of Figure 2. A Gaussian beam at 1550 nm (ORION 1550 nm Laser Module) with the fiber collimator is injected to a half-wave plate and a linear polarizer, which align the polarization according to the requirement of three phase-only reflective SLMs (Holoeye PLUTO-2.1-TELCO-013). Each SLM has 1920×1080 pixels with the pixel pitch of 8μm, serving as a reconfigurable wavefront modulator. SLM0 is employed to split the single incident beam with the radius of 1.63mm into 20 Gaussian beams without overlap as the input DCS mode, and the position distribution in the transverse plane is shown in inset (2) of Figure 2. The beams are arranged in a triangular lattice in order to encode more spins on a single SLM, and the radius of each beam is ~610μm. The selection of the beam radius would be discussed in the section of "Scalability". Both phase patterns for spin-encoding and beam-splitting are applied on SLM1, while the beam-recombining phase pattern is applied on SLM2. SLM1 and SLM2 perform the OVMM of $A\sigma$ together. The splitting ratio of each region on SLM1, which is defined as the ratio of the complex amplitudes of the beams after the splitting, is consistent with the corresponding column of $A$. The recombining ratio of each region on SLM2 is an all-one vector, which sums up the incident optical complex amplitudes. For example, one of the twenty ($N$=20) circular beam-recombining phase pattern is shown in inset (3) of Figure 2. It should be mentioned that the phase pattern on SLM2 only depends on the dimensionality and the position distribution of the DCS mode, *e.g.* 20 beams in a triangular lattice in this work, and keeps constant during the solving process. Moreover, the radius of each beam-splitting and recombining region is set to 1175μm that is almost twice as long as the beam radius to conveniently align each Gaussian beam with $N$=20. Actually, the radius of each circular region could be reduced to about 1.5 times of the Gaussian beam radius according to our estimation while there is no significant overlap between adjacent beam spots. Thus for $N$=30, each beam-splitting/recombining region on SLM is set to 950μm (~1.55×610μm). The detailed discussion about the beam radius is provided in the section of "Scalability". After SLM2, there is a pinhole to filter out the unwanted diffraction components. Before the camera, a lens aligns the beams along the direction of the optical axis. Finally, the DCS mode is detected by the Hamamatsu InGaAs Camera C12741-03. The methods for calibrating the optical system and generating the phase patterns are provided in Supplementary Note 3. Additionally, a CPU is employed to perform the required process in the electronic domain, including the pretreatment of the adjacent matrix, generating the phase patterns on SLMs, flipping the spins, calculating the Hamiltonians and executing other steps required in the adopted heuristic algorithm.

**Ground state search for different Ising models.** To verify our proposed PEIDIA, several models with different complexities and numbers of spins have been experimentally solved. The first model is an antiferromagnetic Möbius-Ladder model with $N=20$ (denoted as model 1), in which the nonzero entries are $J_{ij} = -1$ as illustrated in the inset of Figure 3a. First, a single run of the PEIDIA is conducted, where a final accepted state is obtained after 600 iterations, and the measured experimental Hamiltonian evolution is illustrated in Figure 3a. Figures 3b and 3c show the detected images and the beam intensities of the output fields of the randomly generated initial state and the final accepted state, respectively, and the positions of both states are marked in Figure 3a. In Figures 3b and 3c, the intensity of each beam is represented by the average power of central 9 pixels inferred from the grayscale. In the eigenvalue matrix **D** of the 20-dimensional Möbius-Ladder model, the first 11 eigenvalues are negative while the last 9 eigenvalues are positive. Thus, the beams with number 1-11 are marked as the "negative" beams corresponding to the negative eigenvalues, while the rest are denoted as the "positive" beams in Figures 3b and 3c. In Figure 3b, the intensity mainly concentrates on the "negative" beams, indicating that the initial state is an excited state with a high Hamiltonian ($H=11.5$). Actually, Eq. (4) indicates that the optical intensities are expected to be more concentrated on the "positive" beams to achieve lower Hamiltonians. Figure 3c shows that the intensity finally concentrates on the "positive" beams 19 and 20, while there are almost no signals on the "negative" beams, which corresponds to a low value of the Hamiltonian ($H=-26.0$). The results in Figure 3 indicate that the accepted spin state evolves from an initial state with a high Hamiltonian to a final state with a low Hamiltonian, thus the PEIDIA indeed minimizes the Hamiltonian.

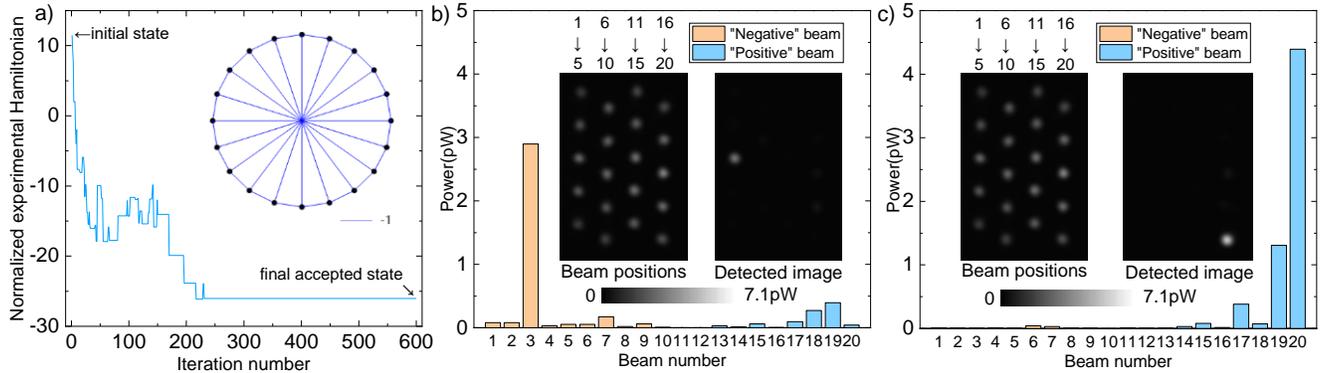

**Figure 3. Results of the ground state search for the 20-dimensional Möbius-Ladder model. a)** An experimentally measured Hamiltonian evolution curve in a single run. The inset shows the Ising model to be solved. Two arrows denote the position of the initial state in **b** and the final accepted state in **c**, respectively. **b)** The beam power of the detected image corresponding to a randomly generated initial state with a high Hamiltonian. The left inset denotes the positions of 20 beams. The color bar denotes the incident power on each pixel. **c)** The beam power of the detected image corresponding to a final accepted state with a low Hamiltonian. The left inset denotes the positions of 20 beams. The color bar denotes the incident power on each pixel.

In the experiment, the PEIDIA has been run for 100 times, and the corresponding Hamiltonian evolutions are depicted in Figure 4a. In each run, the initial state of the spins is randomly generated. Most of the curves converge to the low Hamiltonians within 400 iterations, and the finally obtained Hamiltonians are very close to the ground state Hamiltonian $H=-26$ which is denoted as the black dashed line in Figure 4a. Such distribution may be mainly due to the systematic error and the detection noise. Actually, the target of the PEIDIA is to obtain the spin vector of the ground state, rather than the actual value of the Hamiltonian. Thus, the accepted spin vectors in each iteration corresponding to all curves in Figure 4a are extracted to calculate the theoretical Hamiltonians with Eq. (1), and then the ground state probability is obtained by counting the proportion of the ground state Hamiltonian for each iteration within all 100 runs. The ground state probability versus the iteration number is plotted as the red curve in Figure 4b. It can be seen that as the initial states are randomly generated, the ground state probability is almost 0 in the range of the iteration number less than 50. Then the probability would experience a rapid growth from the iteration number 50 to 300, and gradually converge in the end. The final ground state probability is around 0.99 after 400 iterations, indicating that almost all of the 100 runs can successfully obtain the ground states. For comparison, the algorithm simulation, which does not include the simulation of the entire optical system, has also been carried out for 10000 times on a computer with the same parameters as the experimental settings, and the ground state probability versus the iteration number is plotted as the black curve in Figure 4b. It can be seen that the experimental curve matches very well with the simulation.

As shown in the Figure 4a, the experimental Hamiltonians are distributed around the ground state Hamiltonian in the final stage of searching, indicating that the systematic error and the detection noise cannot be neglected due to the limited performance of the experimental devices. Such error and noise would cause some deviations of the actual transformation matrices, the input vectors and the detected signals from the theoretical expectations. To quantify the influence of these two factors, the parameter of fidelity $f$ is introduced with

$$f = \frac{|\mathbf{I}^T \mathbf{I}_{\text{theo}}|}{|\mathbf{I}||\mathbf{I}_{\text{theo}}|}. \tag{6}$$

In Eq. (6), $\mathbf{I}$ is the intensity vector measured by the camera and $\mathbf{I}_{\text{theo}} = (\mathbf{A}\boldsymbol{\sigma})^* \odot (\mathbf{A}\boldsymbol{\sigma})$ ($\odot$ denotes the element-wise multiplication) is the theoretical output intensity vector, which is calculated by the target transformation matrix $\mathbf{A}$ and the sampled spin vector $\boldsymbol{\sigma}$. Thus, $f$ could evaluate the accuracy of the output intensity vectors, since both the accuracy of the OVMM and the detection noise are involved. According to Eq. (6), $f$ is normalized within $[0, 1]$ and $f = 1$ represents that the two vectors are perfectly parallel to each other. The fidelities of all $6 \times 10^4$ experimental samples are calculated, and the probability distribution of $f$ is illustrated in Figure 4c, and the average value is 0.99978±0.00039, indicating that our transformation scheme is quite accurate.

To present the "on-demand" ability of our proposal, we have considered two other randomly generated and fully connected spin-glass models with spin numbers of 20 and 30 (denoted as model 2 and 3 respectively), in which nonzero entries are uniformly distributed in $J_{ij} \in \{-1, 1\}$. The results of model 2 are shown in Figures 4d-4f, while those of model 3 are shown in Figures 4g-4i. The spin couplings of model 2 and 3 are shown in the insets of Figures 4d and 4g respectively. The measured accepted experimental Hamiltonians for 100 runs are presented in Figures 4d and 4g, where the theoretical ground state Hamiltonians are shown as the black dashed lines ($H$=-62 for model 2 and $H$=-117 for model 3). Here, the number of iterations for each run is increased to 1200 and 2000 respectively since model 2 and 3 have higher graph densities than model 1. It can be seen that most of the curves converge within 600 iterations for model 2 and 1200 iterations for model 3. The ground state probabilities versus the iteration number are also calculated and plotted in Figures 4e and 4h with the final ground state probabilities of 0.97 and 0.85 for model 2 and 3, respectively. Both results indicate that our PEIDIA is capable of solving such complex and fully connected models. Furthermore, the fidelity distributions of the output vectors are also calculated and shown in Figures 4f and 4i respectively. The average fidelities of the sampled output intensity vectors for model 2 and 3 are 0.99976±0.00044 and 0.99942±0.00131 respectively, which are very close to that in model 1. Meanwhile, the fidelities of the corresponding transformation matrices are 0.9989, 0.9994 and 0.9969 for model 1-3, respectively (see Supplementary Note 4). However, for the 30-spin demonstration, the experimental ground state probability becomes a bit lower than that in the simulation. Such result is mainly due to the decreased signal-to-noise ratio of the experimental Hamiltonian with increasing dimensionality $N$ (see detailed discussion of the searching accuracy in Supplementary Note 6).

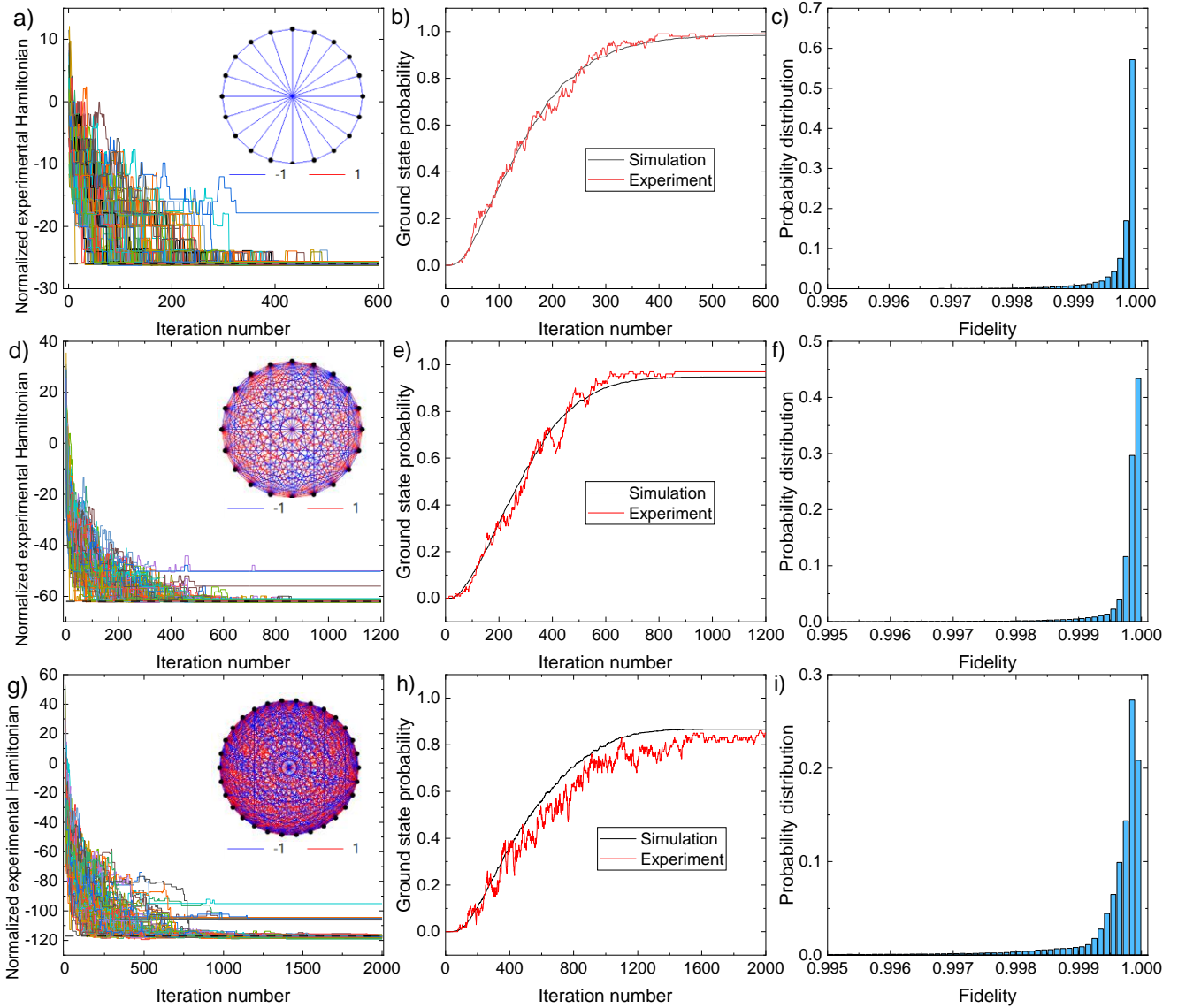

**Figure 4. Results of the three demonstrated models. a)-c)** The 100 experimental Hamiltonian evolution curves for model 1-3 respectively. The normalization method is described in Supplementary Note 5. The inset shows the spin interactions of the models where the red and blue lines denote $J_{ij}=1$ and $-1$ respectively. The black dashed lines indicate the theoretical ground state Hamiltonians. **d)-f)** The ground state probabilities versus the iteration number over 100 runs for model 1-3 respectively. **g)-i)** The fidelity distributions of the sampled output intensity vectors for model 1-3 respectively.

## Discussion

According to our previous work[40,41], each pattern on SLM is the superposition of a series of phase gratings, hence abundant pixels have to be employed to perform such complex pattern with enough accuracy. In this section, we would estimate the upper limit of the number of spins that can be arranged. Firstly, we refer to the Gaussian beam radius function given by

$$w(w_0, \lambda, z) = w_0\sqrt{1 + \left(\frac{\lambda z}{\pi w_0^2}\right)^2}, \tag{7}$$

where $w_0$ is the radius of the beam waist between SLM1 and SLM2, $\lambda$ is the wavelength and $z$ is the axial distance relative to the beam waist. To arrange as many spins as possible for a given $\lambda$ and $z$, it is necessary to ensure that the beam radii on both SLM1 and SLM2 are the same, which can be achieved by locating the beam waist at the midpoint between SLM1 and SLM2. For our experiment, $\lambda$=1550nm and $z_{SLM}$=0.377m (half of the distance between SLM1 and SLM2). By taking the derivative of the Gaussian beam radius function and setting $\partial w/\partial w_0 = 0$, the radius of beam waist should be $w_0$=431μm, which corresponds to a minimum beam radius $w_{SLM}$=610μm on SLM1 and SLM2. Thus, in our experimental setup, the beam radii on SLM1 and SLM2 are fixed and aligned to ~610μm. Additionally, the radius of each circular pattern on SLM1 and SLM2 should be set to larger than $1.5w_{SLM}$ so that the beam overlap is about ~1% (estimated by the overlap integral of two Gaussian modes). Thus, for the 20-spin and 30-spin models, the radius of each beam-splitting/recombining pattern on SLM1 and SLM2 is set to 1175μm and 950μm, respectively. It also should be noted that, due to the paraxial approximation and the Nyquist-Shannon sampling theorem[46], increasing the number of spins would not require larger beam-splitting/recombining regions. Therefore, the minimum radius of the circular pattern is ~950μm in our current experimental setup, and the maximum number of spins is ~36. Another factor that would limit the scalability of the PEIDIA is the noise level of the experimental Hamiltonian (see detailed analysis in Supplementary Note 6). In the experiment of model 1-3, all the noise levels are less than the corresponding minimum Hamiltonian variations, thus high ground state probabilities are achieved. According to our analysis, the signal-to-noise ratio of the experimental Hamiltonian is approximately proportional to $1/\sqrt{N}$. Such result indicates that the searching accuracy of the PEIDIA would deteriorate in high-dimensional conditions, which would lead to the failure of searching the ground state.

In the future work, the number of spins could be increased by reducing the beam radius, adjusting the distances between SLMs properly (the distances should have lower boundaries as paraxial approximation is applied), or utilizing shorter operation wavelength according to Eq. (7). For instance, if the distance between SLM1 and SLM2 is set to 0.4m and $\lambda$=800μm, the minimum beam radii on SLM1 and 2 are ~320μm and the minimum radius of the circular region is ~500μm so that that ~132 beams could be processed on SLM1 and SLM2. Furthermore, if 4K SLMs (3840×2160 pixels) are adopted, the spin number can be increased to ~520 with the same arrangement as our present setup. Besides, the detector with lower noise and higher dynamic range would be helpful to achieve higher searching accuracy of the PEIDIA in high-dimensional conditions.

Although this work is also based on SLMs, the difference between our proposal and Pierangeli et al.'s work[34] relies on the employed OVMM. In the number of the reconfigurable parameters is $2N$, which is contributed by the amplitude modulation and the target intensity pattern. It should be noticed that an Ising interaction matrix without external field has the independent entries of $N(N-1)/2$. Therefore, the OVMM based on a Fourier lens cannot handle arbitrary Ising models. Compared with Pierangeli et al.'s work[34], in which each spin is encoded by a single SLM pixel, our employed OVMM utilized more pixels to form a spin for arbitrary matrix transformations, hence it can solve arbitrary Ising models — that is to say, our demonstration trades the number of implementable spins for the on-demand characteristic.

As mentioned above, our PEIDIA only requires one non-unitary OVMM with proper pretreatment. Besides, the Ising spins are encoded on the phase term of the optical field and only intensity measurement is needed to calculate the Hamiltonian. However, in the on-chip proposal of tunable MZI network, two cascaded Reck schemes are utilized to perform arbitrary OVMM since only unitary matrix transformations can be performed by the Reck scheme. Each Reck scheme requires $N(N-1)/2$ MZIs[37]. For example, a 20-dimensional Reck scheme totally needs 190 MZIs, which consist of 380 beam splitters and 380 phase shifters. Such cascaded structure would impede its high-dimensional implementations. Nevertheless, the primary advantage of the proposal with tunable MZI network is the achievement of the Ising machine on a photonic chip. It should be mentioned that our architecture maybe implemented with integrated photonic devices. As shown in Figure 2, the experimental demonstration mainly consists of lens, SLMs and camera. The functions of SLMs for OVMM and the lens could be realized with tunable metasurfaces[47,48] The spin vector could be updated via high-speed phase modulators[49] instead of refreshing SLMs. Besides, the high-speed photodetectors[50] could be employed to measure the output intensities.

The time cost of our demonstration of the PEIDIA consists of the pretreatment cost in the electronic domain and the iteration cost during the annealing process. In the pretreatment stage, the time complexity of the eigen-decomposition is $O(N^3)$ [44] and the generation of the phase patterns on the SLMs is $O(N^2)$. In fact, the pattern on SLM0 is a beam-splitting pattern and that on SLM2 is a beam-recombining pattern, which could be pre-generated before the annealing process. Different beam-splitting patterns on SLM1 would correspond to different Ising problems, and the generation of each pattern takes about 11min for the 20-spin experiment and 17min for the 30-spin experiment. Such pre-generation could be done while solving the previous problems. During the annealing process, the beam-splitting and recombining patterns on SLM0-2 are unchanged, and only a pre-generated constant phase delay mask encoding the sampled spin state is appended to SLM1 in each iteration, as shown in Figure 1. Therefore, the time cost is primarily determined by the optoelectronic iterations. The time cost per iteration in optical domain $t_o$ depends on the propagation time of light $t_p$, the updating time of SLM1 $t_u$ and the detection time of the camera $t_d$, leading to a total $t_o = t_p + t_u + t_d \approx 0.32s$. The rest operations in the electronic domain are the same as those in the algorithm simulation and the time cost $t_e$ is negligible. Thus, the total time cost per iteration is $t_{iter} = t_o + t_e \approx 0.32s$. In each iteration, the OVMM can perform $F = 2N^2 + 2N$ floating-point operations (FLOPs)[33], including $2N^2$ multiplications of $\mathbf{A}\boldsymbol{\sigma}$ and $2N$ multiplications in the intensity detection process. For instance, for model 3 ($N$=30), the operation speed of the OVMM is $R = F/t_{iter}$ =5.81kFLOP/s. The total energy consumption of the optical system is $P$=16W, including the power of laser and camera (see detailed analysis in Supplementary Note 7). Thus, the energy consumption is $e_{ff} = P/R$ =2.75mJ/FLOP. In the future, the PEIDIA based on integrated devices would achieve more spins, smaller size and less optical time cost, which could lead to higher computation speed and lower energy consumption.

**Conclusions**

In summary, our proposed PEIDIA provides an architecture that can map arbitrary Ising problems to a photonic system. The PEIDIA provides a fully reconfigurable and high-fidelity optical computation, which can accelerate the vector-matrix multiplication with the parallel propagation of light. The PEIDIA only requires one step of OVMM and intensity detection, which makes the architecture more compact and stable. As a proof of principle, two 20-dimensional and one 30-dimensional Ising problems have been successfully solved with high ground state probability of 0.99, 0.97 and 0.85 respectively. In the spatial-photonic implementation of the PEIDIA, the number of spins can be increased by optimizing the experimental conditions, such as employing shorter wavelength, or higher-resolution SLMs. Meanwhile, the utilization of the detector with lower noise level is significant to ensure the high searching accuracy of the PEIDIA in high-dimensional conditions. In our current experimental demonstration of the PEIDIA, the performances are severely limited by the conversion time between optical and electronic signals. We are still undergoing the corresponding work about the integrated scheme of PEIDIA, and we believe that our architecture could be further improved to achieve large-scale on-demand photonic Ising machines.

## Methods

**Generating the phase-only patterns on SLMs.** The phase-only patterns on SLM0-2 to conduct the beam-splitting and recombining operations are iteratively optimized based on a gradient-descent method, rather than simply taking the argument pattern of the superposed weighted blazed gratings in our previous work[40–43]. The loss function depending on complex-valued parameters is defined as the difference between the target field and the field modulated by the pattern to be optimized. The three patterns are generated according to the actual experimental setup. More details of the method are provided in Supplementary Note 3.

**Calibration of SLMs.** Due to various unavoidable systematic errors such as misalignment in the OVMM system, the phase term and the beam-splitting and recombining ratios of all the patterns need to be calibrated. The calibration method is improved from our previous work[42] based on homodyne detection. Such calibration significantly enhanced the fidelity of OVMM and more details are provided in Supplementary Note 3.

**Annealing parameters.** For model 1, the experimental parameters are: initial temperature $(T_0)_{exp}$=3000, steps per temperature stage $n_{step}$=30, stages of temperature $n_{temp}$=20 and annealing factor $\eta$ =0.9. For 100 average values of $K_1$ in Supplementary Figure 6e, which are the Hamiltonian normalization coefficients (see Supplementary Note 5), 100 corresponding initial temperature $(T_0)_{simu}$ are obtained with $(T_0)_{simu} = (T_0)_{exp}/K_1$. Other parameters are the same as those in experiment. For each $(T_0)_{simu}$, 100 simulations are conducted, hence totally 10000 simulations are conducted to obtained the simulation curve of ground state probability in Figure 4b. For model 2/3, the experimental parameter is $(T_0)_{exp}$=2150/1200, $n_{step}$=40/50, $n_{temp}$=30/40 and $\eta$=0.90/0.92, respectively. The process of obtaining the simulation curve of ground state probability in Figures 4e and 4h of the ground state probability is the same as that in model 1. The full operation process is shown in Supplementary Note 2.

## Data availability

The data that support the plots within this paper and other findings of this study are available from the corresponding authors upon reasonable request.

## Code availability

The codes for the generation of the phase patterns on spatial light modulators in this study are available from the corresponding authors upon reasonable request.

## Supplementary information

See Supplementary Information for supporting content.

**Acknowledgments**
This work was supported by the National Key Research and Development Program of China (2018YFB2200402, 2017YFA0303700), the National Natural Science Foundation of China (Grant No. 61875101). This work was also supported by Beijing academy of quantum information science, Beijing National Research Center for Information Science and Technology (BNRist), Frontier Science Center for Quantum Information, Tsinghua University Initiative Scientific Research Program and Huawei Technologies Co., Ltd. The authors would like to thank Mr. Hejia Zhang, Bowen Xia and Ziyue Yang for their valuable discussions and helpful comments.

**Author contributions**
J.O. and X.F. conceived the idea. J.O. designed and performed the simulations, experiments and data analysis. Y.L. contributed significantly to the numerical simulations. Z.M. and D.K. helped to perform the experiments. J.O. and X.F. wrote the manuscript. X.Z., X.D., K.C., F.L. and W.Z. supervised the manuscript. Y.H. supervised, reviewed and edited the manuscript. All authors read and approved the manuscript.

**Competing interests**
The authors declare no competing interests.

Supplementary Information for

# On-demand Photonic Ising Machine with Simplified Hamiltonian Calculation by Phase Encoding and Intensity Detection

Jiayi Ouyang[1], Yuxuan Liao[1], Zhiyao Ma[1], Deyang Kong[1], Xue Feng[1]*, Xiang Zhang[2], Xiaowen Dong[2], Kaiyu Cui[1], Fang Liu[1], Wei Zhang[1], and Yidong Huang[1]

[1]Department of Electronic Engineering, Tsinghua University, Beijing 100084, China
[2]Central Research Institute, Huawei Technologies Co. Ltd., Shenzhen, 518129, China
*Correspondence: x-feng@tsinghua.edu.cn

**This file includes:**

Supplementary Notes 1 to 7

Supplementary Figures 1 to 9

Supplementary Tables 1 to 5

Supplementary References

**Supplementary Note 1: Detailed deduction in the pretreatment stage and the measurement of the experimental Hamiltonian**

At first, we would like to introduce the principle of solving an Ising problem with our proposed Phase Encoding and Intensity Detection Ising Annealer (PEIDIA). in detail. For a given $N$-spin Ising model, we first decompose the adjacent interaction matrix $\mathbf{J}$ to the symmetric matrix $\mathbf{J}_+ = (\mathbf{J} + \mathbf{J}^T)/2$ (the superscript T denotes the transpose) and the anti-symmetric matrix $\mathbf{J}_- = (\mathbf{J} - \mathbf{J}^T)/2$. As $\mathbf{J}_-$ does not contribute to the Hamiltonian, we use $\mathbf{J}$ to denote the symmetric component $\mathbf{J}_+$ for simplicity. Thus, the real symmetric matrix $\mathbf{J}$ can be decomposed with eigen-decomposition[1],

$$\mathbf{J} = \mathbf{Q}^T \mathbf{D} \mathbf{Q}, \tag{S1}$$

where the $\mathbf{D} = \text{diag}(\lambda_1, \lambda_2, \ldots, \lambda_N)$ is the diagonal eigenvalue matrix with $N$ eigenvalues $\lambda_1, \lambda_2, \ldots, \lambda_N$, and $\mathbf{Q}$ is the corresponding normalized orthogonal eigenvector matrix. By introducing $\mathbf{A} = \sqrt{\mathbf{D}} \mathbf{Q}$:

$$\mathbf{J} = \mathbf{Q}^T \sqrt{\mathbf{D}} \sqrt{\mathbf{D}} \mathbf{Q} = \mathbf{A}^T \mathbf{A}. \tag{S2}$$

With such decomposition, the Ising Hamiltonian $H$ can be written as

$$H(\boldsymbol{\sigma}) = -\frac{1}{2} \sum_{1 \leq i,j \leq N} J_{ij} \sigma_i \sigma_j = -\frac{1}{2} \boldsymbol{\sigma}^T \mathbf{J} \boldsymbol{\sigma} = -\frac{1}{2} (\mathbf{A}\boldsymbol{\sigma})^T (\mathbf{A}\boldsymbol{\sigma}), \tag{S3}$$

where $\boldsymbol{\sigma} = [\sigma_1, \sigma_2, \ldots, \sigma_N]^T$ ($\sigma_i \in \{-1,1\}$) denotes the Ising spin vector. In Eq. (S3), the Hamiltonian is the product of the transformed spin vector $\mathbf{A}\boldsymbol{\sigma}$ and its transpose. It should be mentioned that the eigenvalues $\lambda_1, \lambda_2, \ldots, \lambda_N$ may be negative. For convenience, suppose that $\mathbf{J}$ has $k$ negative eigenvalues that satisfies $\lambda_1 \leq \lambda_2 \leq \cdots \leq \lambda_k < 0 \leq \lambda_{k+1} \leq \cdots \leq \lambda_N$. Thus, the square root of the diagonal eigenvalue matrix $\mathbf{D}$ can be expressed as $\sqrt{\mathbf{D}} = \text{diag}(i\sqrt{-\lambda_1}, i\sqrt{-\lambda_2}, \ldots i\sqrt{-\lambda_k}, \sqrt{\lambda_{k+1}}, \ldots, \sqrt{\lambda_N})$. As $\mathbf{Q} = \{q_{ij}\}$ is a real matrix, the transformation matrix $\mathbf{A}$ has the following form

$$\mathbf{A} = \begin{bmatrix} i\sqrt{-\lambda_1}q_{11} & \cdots & i\sqrt{-\lambda_1}q_{k1} & i\sqrt{-\lambda_1}q_{1,k+1} & \cdots & i\sqrt{-\lambda_1}q_{1N} \\ \vdots & \ddots & \vdots & \vdots & \ddots & \vdots \\ i\sqrt{-\lambda_k}q_{k1} & \cdots & i\sqrt{-\lambda_k}q_{k,k} & i\sqrt{-\lambda_k}q_{k,k+1} & \cdots & i\sqrt{-\lambda_k}q_{kN} \\ \sqrt{\lambda_{k+1}}q_{k+1,1} & \cdots & \sqrt{\lambda_{k+1}}q_{k+1,k} & \sqrt{\lambda_{k+1}}q_{k+1,k+1} & \cdots & \sqrt{\lambda_{k+1}}q_{k+1,N} \\ \vdots & \ddots & \vdots & \vdots & \ddots & \vdots \\ \sqrt{\lambda_N}q_{N,1} & \cdots & \sqrt{\lambda_N}q_{N,k} & \sqrt{\lambda_N}q_{N,k+1} & \cdots & \sqrt{\lambda_N}q_{N,N} \end{bmatrix}, \tag{S4}$$

where the first $k$ rows are imaginary, while the rest $(N - k)$ rows are real. Similarly, since the spin vector $\boldsymbol{\sigma}$ is also real, the first $k$ elements of the transformed spin vector $\mathbf{E} = \mathbf{A}\boldsymbol{\sigma}$ are imaginary while the rest elements are real. Thus it can be written as $\mathbf{E} = \mathbf{A}\boldsymbol{\sigma} = [ie_1 \quad \cdots \quad ie_k \quad e_{k+1} \quad \cdots \quad e_N]^T$ ($e_i \in \mathbb{R}$). In the PEIDIA, the optical vector-matrix multiplication (OVMM) is set to $\mathbf{A}$ and the input vector is set to $\boldsymbol{\sigma}$. Since only the intensity of the light field can be detected, the output intensity vector is

$$\mathbf{I} = \mathbf{E}^* \odot \mathbf{E} = [e_1^2 \quad \cdots \quad e_k^2 \quad e_{k+1}^2 \quad \cdots \quad e_N^2]^T = [I_1 \quad \cdots \quad I_k \quad I_{k+1} \quad \cdots \quad I_N]^T, \tag{S5}$$

where $\odot$ denotes element-wise multiplication. Thus Eq. (S3) can be further rewritten as

$$H(\boldsymbol{\sigma}) = -\frac{1}{2} \mathbf{E}^T \mathbf{E} = (I_1 + I_2 + \cdots + I_k - I_{k+1} - \cdots - I_N)/2. \tag{S6}$$

Then the relation between the measured intensities and the corresponding Hamiltonian is

$$H = -\frac{1}{2}\boldsymbol{\sigma}^{\mathrm{T}}\mathbf{J}\boldsymbol{\sigma} = \frac{1}{2}\left(\sum_{i,\,\lambda_i<0} I_i - \sum_{i,\,\lambda_i>0} I_i\right). \tag{S7}$$

In conclusion, in the pretreatment stage, the required optical transformation matrix **A** is obtained by the eigen-decomposition of the Ising interaction matrix **J**. The experimental Hamiltonian can be calculated from the output intensities of the OVMM. In the main text, a 20-spin antiferromagnetic Möbius-Ladder model (model 1) and two randomly generated, fully connected spin-glass models (model 2 and 3 with 20 and 30 spins respectively) are demonstrated, and their interaction matrices **J** and corresponding transformation matrices **A** are shown in Supplementary Figure 1 respectively.

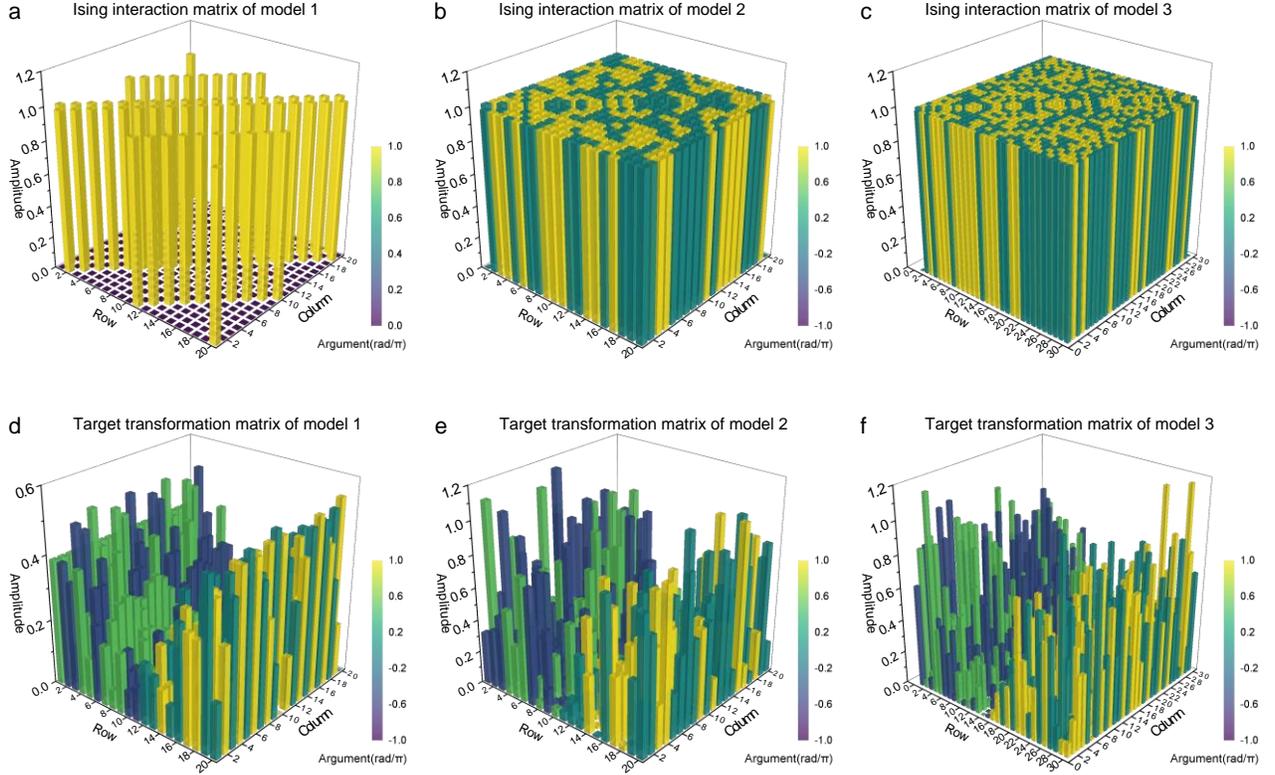

**Supplementary Figure 1.** Related matrices of model 1-3. (a)-(c) The Ising interaction matrices of model 1-3 respectively. (d)-(f) The target transformation matrices **A** of model 1-3 respectively.

**Supplementary Note 2: Full operation process of the PEIDIA**

The operation principle of PEIDIA is modified from the simulated annealing algorithm[2,3] and inspired by the fast simulated annealing algorithm[4]. The process of each run is:

**Input:** Ising interaction matrix **J**, number of spins $N$, transformation matrix **A**, initial random spin vector $\boldsymbol{\sigma}^{(0)}$, initial Hamiltonian $H^{(0)}$, initial annealing temperature $T_0$, steps per temperature $n_{\mathrm{step}}$, stages of the temperature $n_{\mathrm{temp}}$, annealing coefficient $\eta$, Cauchy scaling coefficient $\alpha$.

**Output:** $\boldsymbol{\sigma}$

Initialization: annealing temperature: $T \leftarrow T_0$, Hamiltonian: $H \leftarrow +\infty$, spin vector: $\boldsymbol{\sigma} \leftarrow \boldsymbol{\sigma}^{(0)}$, patterns on SLMs: calibrated patterns corresponding to **A**.

For $i = 1$ to $n_{\text{temp}}$ do

    For $j = 1$ to $n_{\text{step}}$ do

        Generate a Cauchy random variable $x \sim C(x; 0, \alpha T)$ until $m = \text{round}(|x|) < N$

        If $m = 0$

            $m \leftarrow 1$

        Else if $m > N/2$

            $m \leftarrow N - m$

        End if

        Generate the spin state $\boldsymbol{\sigma}_{\text{next}}$ by randomly flip $m$ spins from $\boldsymbol{\sigma}$

        Update SLM1 according to $\boldsymbol{\sigma}_{\text{next}}$

        Calculate $H_{\text{next}}$ according to the measured beam intensities

        If $H_{\text{next}} < H$ or $\text{rand}(0,1) < \exp\left(-\frac{H_{\text{next}} - H}{T}\right)$

            $\boldsymbol{\sigma} \leftarrow \boldsymbol{\sigma}_{\text{next}}$

            $H \leftarrow H_{\text{next}}$

        End if

    End for

    $T \leftarrow \eta T$

End for

Return $\boldsymbol{\sigma}$

**Supplementary Note 3: Experimental demonstration of the PEIDIA**

    The experimental setup is illustrated in Supplementary Figure 2. The whole system is placed on an optical table with vibration isolators and covered by a glass cabinet while the temperature of the laboratory is also kept constant. The operating wavelength of the laser (ORION 1550nm Laser Module) is 1550nm. Each employed reflective phase-only SLM (HOLOEYE PLUTO-2.1-TELCO-013) has 1920×1080 pixels with the pixel pitch of 8μm. The distance between SLM0 and SLM1 is $L_{01}$=0.820m. The distance between SLM1 and SLM2 is $L_{12}$=0.754m. The distance between SLM2 and pinhole is $L_{2p}$=0.373m. The final output field is measured by the Hamamatsu InGaAs C12741-03 camera. To weaken the influence of the detection noise, the measured image of each sampled state is the average of 3 adjacent frames. Supplementary Figure 3 shows the experimentally utilized SLM patterns for model 2 as an example. Supplementary Figure 3a is the beam-splitting pattern on SLM0. 20 beam-splitting and beam-recombining circular regions are shown in Supplementary Figures 3b and 3c respectively with the radius of each region is 1175μm. Supplementary Figure 4 shows the SLM1 pattern used in the experiment of model 3, which contains 30 beam-splitting regions with radius of 950μm.

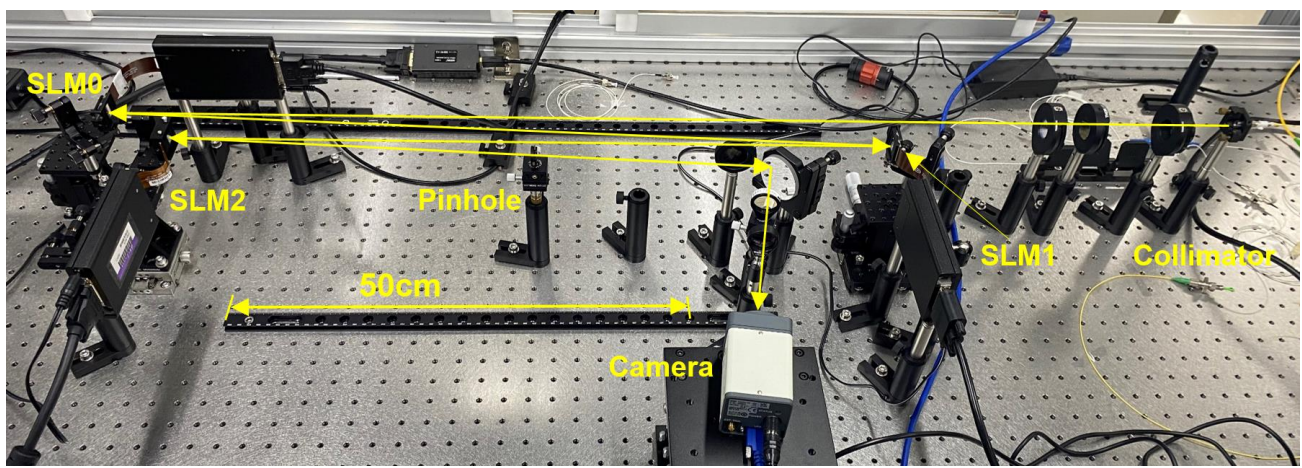

**Supplementary Figure 2**. The experimental setup.

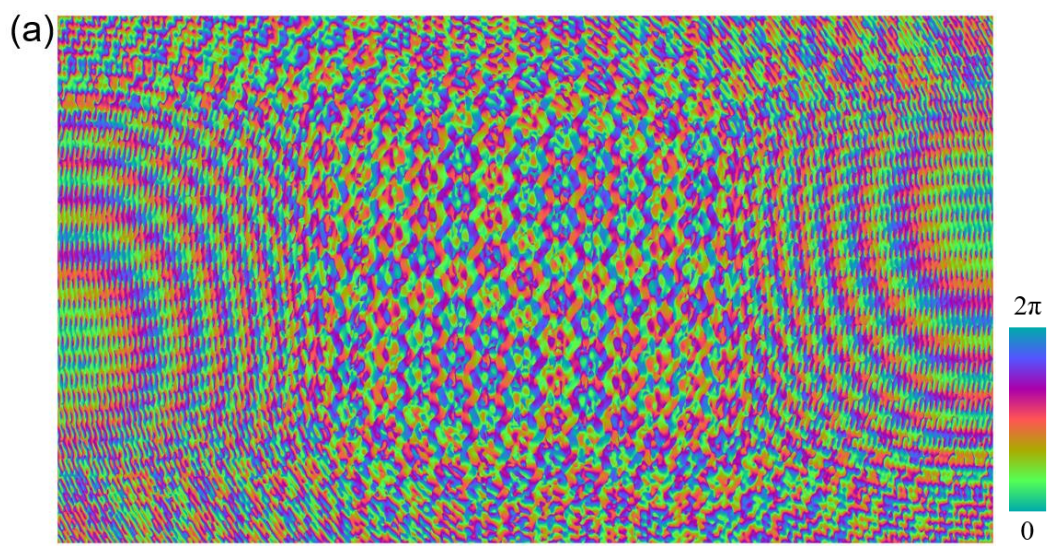

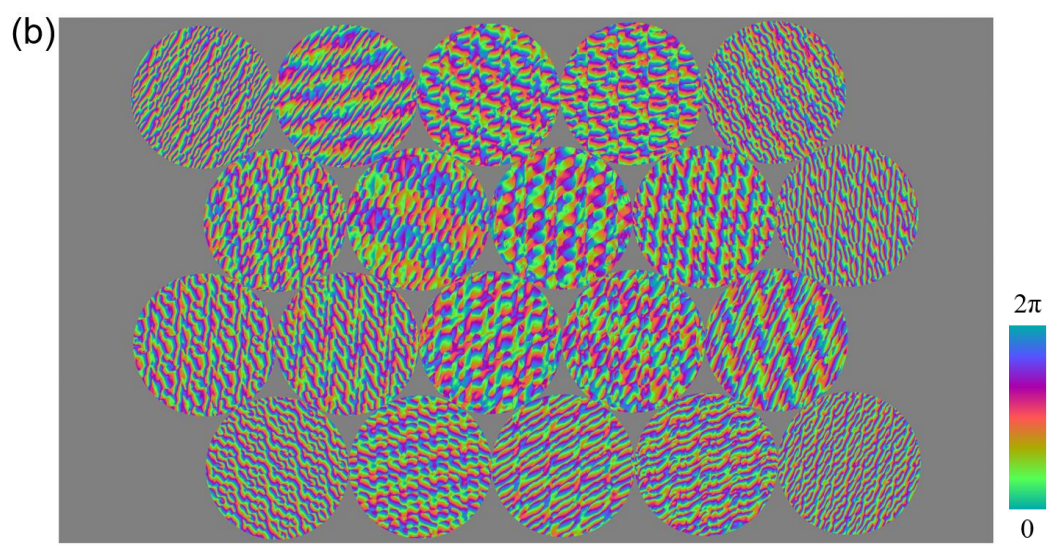

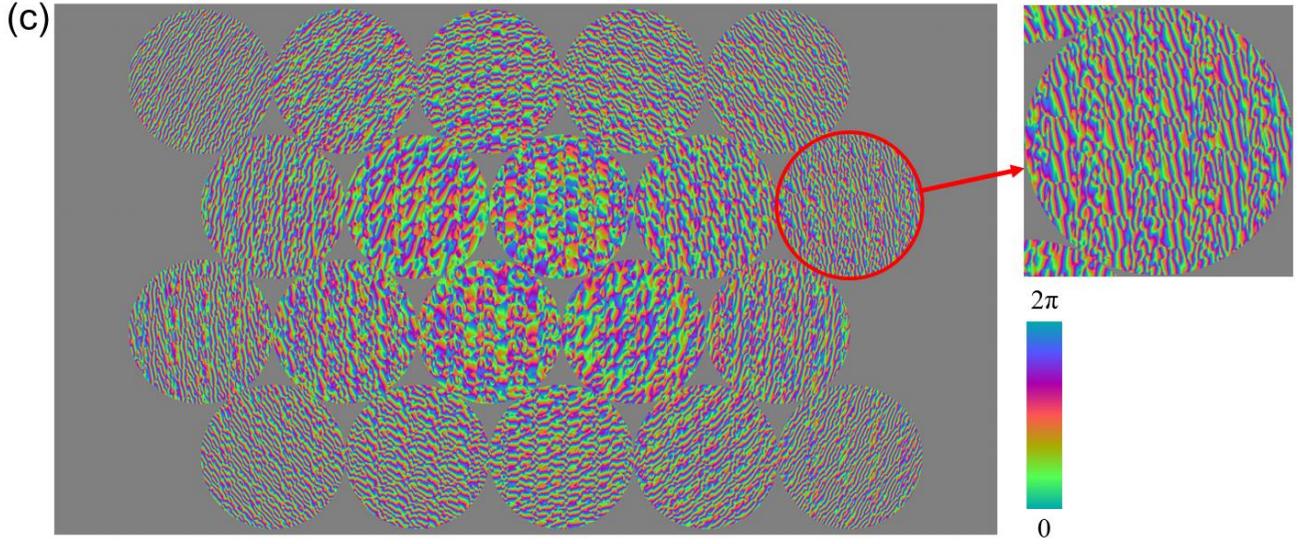

**Supplementary Figure 3**. The experimentally utilized phase patterns on spatial light modulators (SLMs) for model 2 (not superposed with the global blazed grating). (a) The beam-splitting pattern on SLM0. (b) The beam-splitting pattern on SLM1. (c) The beam-recombining pattern on SLM2. The inset shows a detailed beam-recombining region denoted by the red circle. The color bar indicates the phase value.

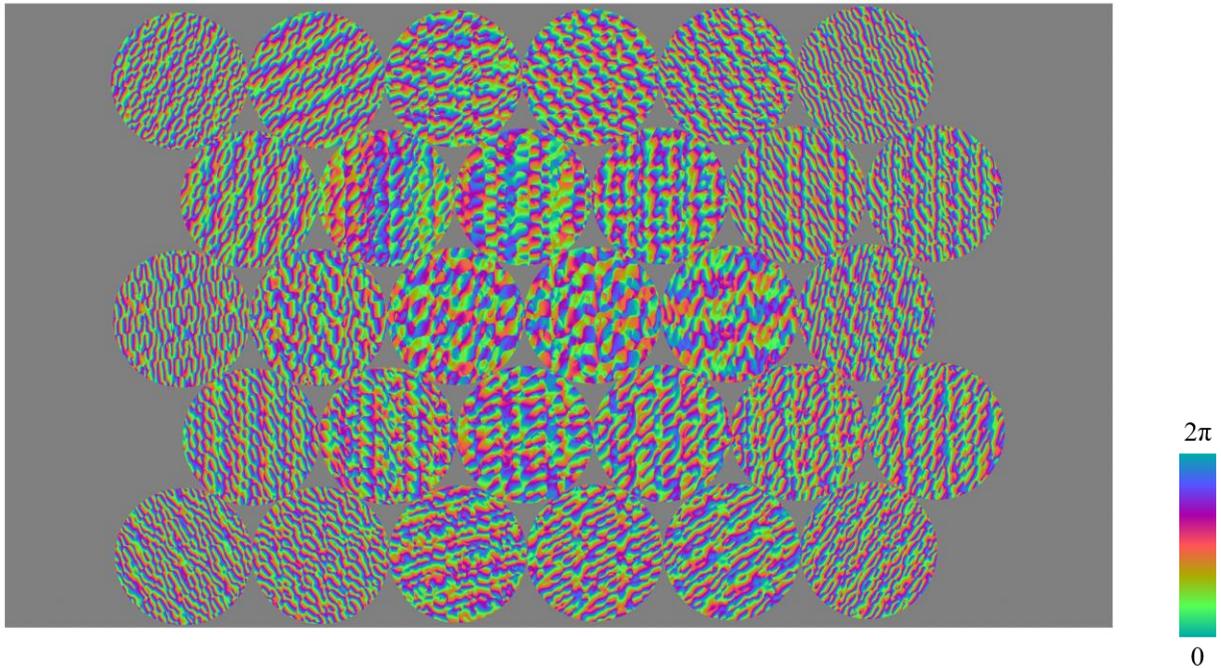

**Supplementary Figure 4**. The experimentally utilized beam-splitting phase pattern for model 3 (not superposed with the global blazed grating). The color bar indicates the phase value.

**The method of generating phase patterns on SLMs.** In our previous work[5,6], the method of generating the phase patterns on SLMs has been presented and the main procedure are described as follows. As mentioned in the main text, three SLMs are employed to generate the required beams, indexed SLM0, SLM1 and SLM2 respectively. SLM0 splits the incident beam into $N$ beams, while SLM1 and SLM2 split and interweave the beams to perform multiplication and addition operations. The pixels outside the circular regions are filled with alternating 0 and $\pi$ phase modulation according to the checkerboard method. Besides, as the employed SLMs are reflective, an additional global blazed grating with the period of 4 pixels is applied to each SLM pattern in order to concentrate the optical intensity in the first diffraction order. The beam from the collimator is centered at coordinate [0, 0] in the transverse $xy$ plane. Then the beam is split and projected to $N$ different circular regions on SLM1, as shown in Supplementary Figures 3b and 4. The ideal modulation function of SLM0 is the superposition of a series of blazed gratings[5,6]:

$$H_{\text{SLM0,ideal}}(\mathbf{r}) = \sum_{n}^{N} \alpha_n \exp\left(\frac{-\mathrm{i}k}{L_{01}} \mathbf{r}_n \cdot \mathbf{r}\right), \tag{S8}$$

where $\mathbf{r}$ is the position vector in the $xy$ plane, $k$ is the wavenumber, $\mathbf{r}_n$ is the position vector of the center of the $n$th region of SLM1, $\alpha_n$ is the element of the input vector.

The ideal modulation function of $n$th circular region on SLM1 comprises two parts: a spin-encoding pattern and a beam-splitting pattern. The spin-encoding pattern just applies a phase delay of 0 or $\pi$ to the beam according to its corresponding spin $\sigma_n = \pm 1$ and the beam-splitting pattern is a superposition of a lens mask and $N$ blazed gratings[5,6]:

$$H_{\text{SLM1,ideal},n}(\mathbf{r}) = \sigma_n \sum_{m=1}^{N} \beta_{mn} \exp\left[-\mathrm{i}k\left(\frac{\mathbf{R}_m - \mathbf{r}_n}{L_{12}} - \frac{\mathbf{r}_n}{L_{01}}\right) \cdot (\mathbf{r} - \mathbf{r}_n) + \theta_{mn}\right] \\ \cdot H_{\text{lens}}(\mathbf{r} - \mathbf{r}_n, f_1), \tag{S9}$$

$$H_{\text{lens}}(\mathbf{r} - \mathbf{r}_n, f_1) = \exp\left(\mathrm{i}\frac{k|\mathbf{r} - \mathbf{r}_n|^2}{2f_1}\right), \tag{S10}$$

where $\beta_{mn}$ is the element of the beam-splitting matrix, $\mathbf{R}_m$ is the position vector of the center of the $m$th region of SLM2, $\theta_{mn}$ is the phase compensation term due to different optical path, and $H_{\text{lens}}$ is the phase modulation function of a lens. The lens with the focus $f_1$ adjusts the location of the beam waist to the middle of SLM1 and SLM2.

Each region on SLM2 recombines the $N$ beams deflected by each region on SLM1 and a pinhole is set to filter out undesired light. The ideal modulation function of $m$th region on SLM2 is[5,6]

$$H_{\text{SLM2,ideal},m}(\mathbf{r}) = \sum_{n=1}^{N} \gamma_{mn} \exp\left[-\mathrm{i}k\left(\frac{-\mathbf{r}_n}{L_{2p}} - \frac{\mathbf{R}_m - \mathbf{r}_n}{L_{12}}\right) \cdot (\mathbf{r} - \mathbf{R}_m)\right] \\ \cdot H_{\text{lens}}(\mathbf{r} - \mathbf{R}_m, f_2), \tag{S11}$$

where $\gamma_{mn}$ is the element of the beam-recombining matrix. The lens with the focus $f_2$ adjusts the location of the beam waist to the pinhole.

There is an evident caveat, however, that the ideal modulation functions above are not pure-phase, thus unattainable with the SLMs employed in our experiments. To address this issue, our previous work maps the value to a point on unit circle with the same argument[5,6]:

$$H_{\text{SLM}}(\mathbf{r}) = \exp\{\mathrm{i}\arg[H_{\text{SLM,ideal}}(\mathbf{r})]\} \tag{S12}$$

for it is the closest pure-phase solution to the original function in the sense of least squares. It is worth pointing out that this practice is empirical, and the errors for both the light field and the matrix elements would increase drastically with the dimensionality.

In this work, an improved iterative algorithm is employed to obtain the optimal phase-only patterns, which produce the light field distribution on the target plane approximating to an ideal distribution. The fundamental theory here is gradient descent. The main issue is how to optimize a real criterion function depending on complex-valued parameters. According to ref. [7], the optimizing theorems are handy to us.

(i)   Preparations

The first step of optimization is to discretize the problem so that the computer processing can be performed. Consider a continuous model of light field propagation: light field given two parallel planes with aligned transverse coordinates, the propagation of the light field $u(x, y, 0)$ on the first plane to the field $u(x', y', z_0)$ on the second plane can be described as[8]

$$u(x', y', z_0) = -u(x, y, 0) \star \frac{1}{2\pi} \frac{\partial}{\partial z'} \left[ \frac{e^{-ikr}}{r} \right]_{z=z_0} = \text{IFT} \left\{ \text{FT}\{u(x,y,0)\} \times \text{FT} \left\{ -\frac{1}{2\pi} \frac{\partial}{\partial z'} \left[ \frac{e^{-ikr}}{r} \right]_{z=z_0} \right\} \right\}, \tag{S13}$$

where $r = \sqrt{(x-x')^2 + (y-y')^2 + z^2}$, $\star$ denotes the convolution, and Fourier transform (FT) is utilized to compute the convolution.

Then the fields are sampled, and the Fourier transform and convolution are replaced with their discrete counterparts. For the sake of simplicity, the light fields are uniformly sampled in a rectangular area, aligned with both axes, and centered on the origin.

Let $\mathbf{H}_{z_0} \in \mathbb{C}^{3L_x \times 3L_y}$ denote the 2D-DFT of the respond $-\frac{1}{2\pi} \frac{\partial}{\partial z'} \left[ \frac{e^{-ikr}}{r} \right]_{z=z_0}$, and $\mathbf{U}, \mathbf{U}' \in \mathbb{C}^{L_x \times L_y}$ denote the sampling of light fields at $z = 0$ and $z = z_0$, respectively. Noting the non-overlapping condition and the alignment of discrete convolution, the response has to be sampled from an area of double length in each dimension, then applied with $3L_x \times 3L_y$ 2D-DFT and finally clips the result to match the shape of $\mathbf{U}'$. To formulate it:

$$\mathbf{U}' = \text{clip} \left( \text{IDFT}_{3L_x, 3L_y} \left( \text{DFT}_{3L_x, 3L_y}(\mathbf{U}) \odot \mathbf{H}_{z_0} \right) \Delta x \Delta y \right), \tag{S14}$$

where $\odot$ denotes element-wise multiplication, $\Delta x$ and $\Delta y$ denote the sampling interval along each axis, and $\text{clip}(\cdot)$ clips extra zeros from the result matrix.

Now consider a surface at $z = 0$, such as an SLM, and assume that each pixel independently produces a wavefront (rectangular, for example) proportional to their complex modulation coefficient. Let $\mathbf{M} \in \mathbb{C}^{3L_x \times 3L_y}$ denotes the 2D-DFT of the wavefront and $\mathbf{P} \in \mathbb{C}^{L_x \times L_y}$ denotes the coefficient matrix of the hologram. Now

$$\mathbf{U}' = \text{clip} \left( \text{IDFT}_{3L_x, 3L_y} \left( \text{DFT}_{3L_x, 3L_y}(\mathbf{P} \odot \mathbf{U}) \odot \mathbf{H}_{z_0} \odot \mathbf{M} \right) \Delta x \Delta y \right) = \mathbf{T}_{z_0}(\mathbf{P} \odot \mathbf{U}). \tag{S15}$$

Note that the composited operations are defined as $\mathbf{T}_{z_0}(\cdot)$, which is the discrete form of light field propagation function we desire.

(ii)   Generate target light field

After discretization, a target light field is needed for each SLM to be optimized. A convenient approach is to treat the light field produced by ideal patterns as target. As well the initial values of optimization are such set with ideal patterns. To formulate it:

$$\mathbf{U}_\text{T} = \mathbf{T}_{z_0}(H_{\text{SLM,ideal}} \odot \mathbf{U}_\text{I}), \tag{S16}$$

$$\mathbf{P}_0 = H_{\text{SLM,ideal}}. \tag{S17}$$

(iii) Optimization steps

Given a target light field $\mathbf{U}_T$ at $z = z_0$, an incident light field $\mathbf{U}_I$ at $z = 0$, and a hyperparameter $\delta$, optimize phase pattern $\mathbf{P}$ in the sense (Huber Loss):

$$\operatorname{argmin}_{\mathbf{P}} \mathcal{L}(\mathbf{P}; \mathbf{U}_T, \mathbf{U}_I, z_0, \delta) = \sum_{i,j} \begin{cases} \frac{1}{2}\left|\mathbf{T}_{z_0}\left(\frac{\mathbf{P}}{|\mathbf{P}|} \odot \mathbf{U}_I\right) - \mathbf{U}_T\right|_{ij}^2, & \left|\mathbf{T}_{z_0}\left(\frac{\mathbf{P}}{|\mathbf{P}|} \odot \mathbf{U}_I\right) - \mathbf{U}_T\right|_{i,j} \leq \delta \\ \delta \left|\mathbf{T}_{z_0}\left(\frac{\mathbf{P}}{|\mathbf{P}|} \odot \mathbf{U}_I\right) - \mathbf{U}_T\right|_{i,j} - \frac{1}{2}\delta^2, & \text{otherwise} \end{cases}$$

$$= l\left(\mathbf{T}_{z_0}(\mathbf{N}(\mathbf{P}))\right), \tag{S18}$$

where $\mathbf{T}_z(\cdot)$ is previously defined, $\{\cdot$ selects from two values and $\sum \cdot$ sums them up. Other operations here are all element-wise. Note that $l(\cdot)$ contracts the loss function and $\mathbf{N}(\mathbf{P}) = \frac{\mathbf{P}}{|\mathbf{P}|}$ gives a phase-only hologram, which is what we eventually need.

Following the procedures described in ref. [7], we finally derive the gradient of $\mathcal{L}$ with respect to $\mathbf{P}$, which is:

$$\frac{\partial}{\partial \mathbf{P}^*}\mathcal{L} = \frac{1}{2}\mathbf{U}_I \odot \left(-\frac{|\mathbf{P}|}{(\mathbf{P}^*)^2} \odot \mathbf{T}'_{z_0}{}^*\left(\frac{\partial}{\partial \mathbf{T}^*_{z_0}}l\right) + \frac{1}{|\mathbf{P}|} \odot \mathbf{T}'_{z_0}\left(\frac{\partial}{\partial \mathbf{T}^*_{z_0}}l\right)\right), \tag{S19}$$

where

$$\mathbf{T}'_{z_0}(\mathbf{Z}) = \operatorname{clip}\left(\operatorname{IDFT}_{3L_x, 3L_y}\left(\operatorname{DFT}_{3L_x, 3L_y}(\mathbf{Z}) \odot \mathbf{H}^*_{z_0} \odot \mathbf{M}^*\right) \Delta x \Delta y\right). \tag{S20}$$

Finally, the gradient-descent-based algorithm is conducted to optimize the target phase-only hologram. Here is a sketch of the algorithm:

Input: target light field: $\mathbf{U}_T$, incident light field: $\mathbf{U}_I$, hyperparameter: $\delta$, initial phase pattern: $\mathbf{P}$, iteration limit iter.

Output: $\mathbf{P}_o$.

Initialization: optimizer: $\mathcal{O} \leftarrow \mathcal{O}_{\text{RAdam}}$ (Rectified Adam Optimizer[9] is employed in our demonstration), loss function $l$: Huber loss.

$\mathbf{P}_0 = \mathbf{P}$

    For $i = 1$ to iter do

        Compute $\frac{\partial}{\partial \mathbf{P}^*_{i-1}}\mathcal{L}$

        $\mathbf{P}_i = \mathcal{O}.\operatorname{step}\left(\frac{\partial}{\partial \mathbf{P}^*_{i-1}}\mathcal{L}\right)$

    End for

Return $\mathbf{P}_o = \frac{\mathbf{P}_T}{|\mathbf{P}_T|}$

**Calibration method of the OVMM.** Before the experiments, the whole system should be calibrated to achieve high fidelity of the implemented matrix transformation to reduce the systematic error. In the calibration stage, both the phase and amplitude terms are considered. The phase calibration method is improved from that used in our previous work[10].

**Phase calibration.** First, as mentioned above, the phase patterns of SLM1 and SLM2 are generated and set according to the all-one matrix. To measure the phase difference among the elements of the beam-splitting matrix, four output intensity matrices $\mathbf{M}_1$, $\mathbf{M}_2$, $\mathbf{M}_3$ and $\mathbf{M}_4$ have to be measured. In this section, each column of the input matrix would serve as the input vector one by one and the output matrix consists of the corresponding output vectors.

At first, the input matrix $\mathbf{m}_{in}$ of SLM0 is defined as an $N \times (N-1)$ matrix with each column including two non-zero elements 1:

$$\mathbf{m}_{in} = \begin{bmatrix} 1 & 1 & 1 & \cdots & 1 \\ 1 & 0 & 0 & \cdots & 0 \\ 0 & 1 & 0 & \cdots & 0 \\ 0 & 0 & 1 & \cdots & 0 \\ \vdots & \vdots & \vdots & \ddots & \vdots \\ 0 & 0 & 0 & \cdots & 1 \end{bmatrix}. \tag{S21}$$

The process of phase calibration can be described as follows:

Initialization: input matrix: $\mathbf{m}_{in}$, matrix on SLM1: all-one matrix, matrix on SLM2: all-one matrix.

For i = 1 to N − 1 do

    Update SLM0 according to the column i of $\mathbf{m}_{in}$

    Measure the column i output intensity matrix $\mathbf{M}_1$

    Deactivate the region $i + 1$ of SLM1

    Measure the column i output intensity matrix $\mathbf{M}_3$

    Reset the SLM1 pattern

    Deactivate the region 1 of SLM1

    Measure the column i output intensity matrix $\mathbf{M}_4$

    Reset the SLM1 pattern

    Add the phase delay of $\pi/2$ to the region $i + 1$ of SLM1

    Measure the column i output intensity matrix $\mathbf{M}_2$

End for

Here, the operation "deactivate" refers to replacing a specific region with a "mirror" pattern where the phase modulation is uniform. Such pattern can reflect the incident beam away from the next SLM. Then how to use these matrices to conduct the phase calibration is introduced.

According to the process above, the interference intensities between the first and one of the rest of the beams have been measured as $\mathbf{M}_1$. Since none of the SLM patterns are calibrated, the actual input matrix can only be written as

$$\mathbf{m}'_{\text{in}} = \begin{bmatrix} b_{11} & b_{12} & b_{13} & \cdots & b_{1,N-1} \\ b_{21} & 0 & 0 & \cdots & 0 \\ 0 & b_{32} & 0 & \cdots & 0 \\ 0 & 0 & b_{43} & \cdots & 0 \\ \vdots & \vdots & \vdots & \ddots & \vdots \\ 0 & 0 & 0 & \cdots & b_{N,N-1} \end{bmatrix} \tag{S22}$$

The actual transformation matrix is denoted as $A'_{ij} = a_{ij} \exp(i\varphi_{ij})$ so that the output amplitude matrix $\mathbf{m}_1 = \mathbf{A}'\mathbf{m}'_{\text{in}}$ is

$$(m_1)_{ij} = a_{i1}b_{1j}\exp(i\varphi_{i1}) + a_{i,j+1}b_{j+1,j}\exp(i\varphi_{i,j+1}) \tag{S23}$$

and the corresponding measured output intensity matrix $\mathbf{M}_1$ is

$$(M_1)_{ij} = (m_1)^*_{ij}(m_1)_{ij} = a_{i1}^2 b_{1j}^2 + a_{i,j+1}^2 b_{j+1,j}^2 + 2a_{i1}b_{1j}a_{i,j+1}b_{j+1,j}\cos(\varphi_{i1} - \varphi_{i,j+1}). \tag{S24}$$

Similarly, the output amplitude matrix $\mathbf{m}_2$ is

$$(m_2)_{ij} = a_{i1}b_{1j}\exp(i\varphi_{i1}) + a_{i,j+1}b_{j+1,j}\exp[i(\varphi_{i,j+1} + \pi/2)] \tag{S25}$$

and the corresponding measured output intensity matrix $\mathbf{M}_2$ is

$$(M_2)_{ij} = (m_2)^*_{ij}(m_2)_{ij} = a_{i1}^2 b_{1j}^2 + a_{i,j+1}^2 b_{j+1,j}^2 + 2a_{i1}b_{1j}a_{i,j+1}b_{j+1,j}\sin(\varphi_{i1} - \varphi_{i,j+1}). \tag{S26}$$

Obviously, $(M_3)_{ij} = a_{i1}^2 b_{1j}^2$ and $(M_4)_{ij} = a_{i,j+1}^2 b_{j+1,j}^2$.

Therefore, the actual phase difference could be calculated according to

$$\Delta\varphi_{ij} = \varphi_{i1} - \varphi_{i,j+1} = \begin{cases} \arccos\left[\dfrac{(M_1)_{ij} - (M_3)_{ij} - (M_4)_{ij}}{2\sqrt{(M_3)_{ij}(M_4)_{ij}}}\right], & (M_2)_{ij} - (M_3)_{ij} - (M_4)_{ij} \geq 0 \\ -\arccos\left[\dfrac{(M_1)_{ij} - (M_3)_{ij} - (M_4)_{ij}}{2\sqrt{(M_3)_{ij}(M_4)_{ij}}}\right], & (M_2)_{ij} - (M_3)_{ij} - (M_4)_{ij} < 0 \end{cases}. \tag{S27}$$

It should be mentioned that Eq. (S27) can only calculate the phase differences among the elements in the same row. But this does not matter since the phase difference among different columns will not affect the output optical intensities. Finally, the phase-calibration matrix $\Delta\boldsymbol{\varphi}$ is obtained, and the input beam-splitting matrix of SLM1 is modified to $A_{ij}\exp(i\Delta\varphi_{ij})$. It should be mentioned that such $\Delta\boldsymbol{\varphi}$ is independent to the transformation matrix and only determined by the optical setup so that it could be directly applied on other transformations.

**Amplitude calibration.** The process of the calibration of SLM1 can be described as follows.

Initialization: Input matrix: identity matrix, target transformation matrix: $\mathbf{A}$, beam-splitting matrix on SLM1: $(A_{\text{SLM1}})_{ij} \leftarrow A_{ij}\exp(i\Delta\varphi_{ij})$, pattern on SLM2: the same as that in phase calibration stage, repetition times: $n_{\text{amp},1}$.

For $i_{\text{rep}} = 1$ to $n_{\text{amp},1}$ do

    Measure the output intensity matrix $\mathbf{C}$

    $(A_{\text{SLM1}})_{ij} \leftarrow (A_{\text{SLM1}})_{ij}|A_{ij}|/\sqrt{C_{ij}}$

    Generate and update the new phase pattern on SLM1 pattern according to $\mathbf{A}_{\text{SLM1}}$

End for

In the 20/30-dimensional transformation, $n_{\text{amp},1} = 3\sim4$ is enough to obtain high fidelity. It should be mentioned that due to the phase-only operation, the SLM1 could only split the incident beams. Thus, the input vector of SLM0 has to be modified to compensate the difference between the beam-splitting matrix of SLM1 and the target matrix. Here, the actual contribution of each split beam from SLM0 to the output mode has to be measured. This can be achieved by activating each region of the calibrated SLM1 pattern successively (other regions are "deactivated" as mentioned above). The process is described as follows.

Initialization: target transformation matrix: **A**, input vector of SLM0: $(E_{\text{in}})_j \leftarrow \max_i(|A_{ij}|)/\sqrt{\max_i(C_{ij})}$ (**C** is the last output intensity matrix in the amplitude calibration of SLM1), pattern on SLM1: calibrated pattern, SLM2 pattern: the same as that in phase calibration stage, repetition times $n_{\text{amp},2}$.

For $i_{\text{rep}} = 1$ to $n_{\text{amp},2}$ do

    For $j = 1$ to $N$ do

        Deactivate all regions except region $j$ of SLM1

        Measure the output intensity vector $\mathbf{F}_j$

$(E_{\text{in}})_j \leftarrow (E_{\text{in}})_j \max_i(|A_{ij}|)/\sqrt{\max_i(F_j)_i}.$

    End for

    Generate and update the new SLM0 pattern according to $\mathbf{E}_{\text{in}}$

End for

In the 20/30-dimensional condition, $n_{\text{amp},2} = 3\sim4$ is enough to obtain high fidelity.

In the end of this section, the 20-dimensional Discrete Fourier Transform (DFT) matrix is set as the target transformation to evaluate the calibration accuracy:

$$\mathbf{W} = \frac{1}{\sqrt{N}}\begin{bmatrix} 1 & 1 & 1 & \cdots & 1 \\ 1 & \omega & \omega^2 & \cdots & \omega^{N-1} \\ 1 & \omega^2 & \omega^4 & \cdots & \omega^{2(N-1)} \\ \vdots & \vdots & \vdots & \ddots & \vdots \\ 1 & \omega^{N-1} & \omega^{2(N-1)} & \cdots & \omega^{(N-1)(N-1)} \end{bmatrix}, \omega = e^{-i\frac{2\pi}{N}}. \tag{S28}$$

Since the elements of both the DFT matrix and its inverse matrix have the same norms and only differ in arguments, it is very sensitive to the phase term. After the amplitude calibration of SLM1, the output intensity matrix corresponding to the input identity matrix is shown in Supplementary Figure 5a. It can be seen that each beam is equally split by SLM1, but the intensities in different columns are different. Then after the amplitude calibration of SLM0, the matrix **F** is shown as Supplementary Figure 5b, where the matrix elements are much closer to each other than that in Supplementary Figure 5a.

    In order to evaluate the accuracy of the calibration, the input matrix is set as the inverse matrix of the DFT matrix, which is also the conjugate transpose. The output matrix should be the 20-dimensional identity matrix. Obviously, the input vector can be simply achieved by adding phase masks to each SLM1 region according to the arguments of the elements of the target input vector. The measured images are shown in Supplementary Figure 5d, where only one bright spot at different locations is observed in each image with similar intensities and high contrast. The output intensity matrix is shown as

Supplementary Figure 5c where the intensity concentrates on the main diagonal elements. Here the fidelity $f_{\text{mat}}$ is employed to evaluate the accuracy of the matrix transformation:

$$f_{\text{mat}} = \frac{\sum_{ij}(I_{\text{t}})_{ij}(I_{\text{e}})_{ij}}{\sqrt{\sum_{ij}(I_{\text{t}})_{ij}^2}\sqrt{\sum_{ij}(I_{\text{e}})_{ij}^2}}, f_{\text{mat}} \in [0,1] \tag{S29}$$

where $\mathbf{I}_{\text{t}}$ is the theoretical output intensity matrix and $\mathbf{I}_{\text{e}}$ is the experimentally measured intensity matrix. $f_{\text{mat}}$ is normalized within [0,1]. In this demonstration, $f_{\text{mat}}$ is calculated from the output intensity matrix in Supplementary Figure 5c and the identity matrix, resulting in a value as high as 0.99994.

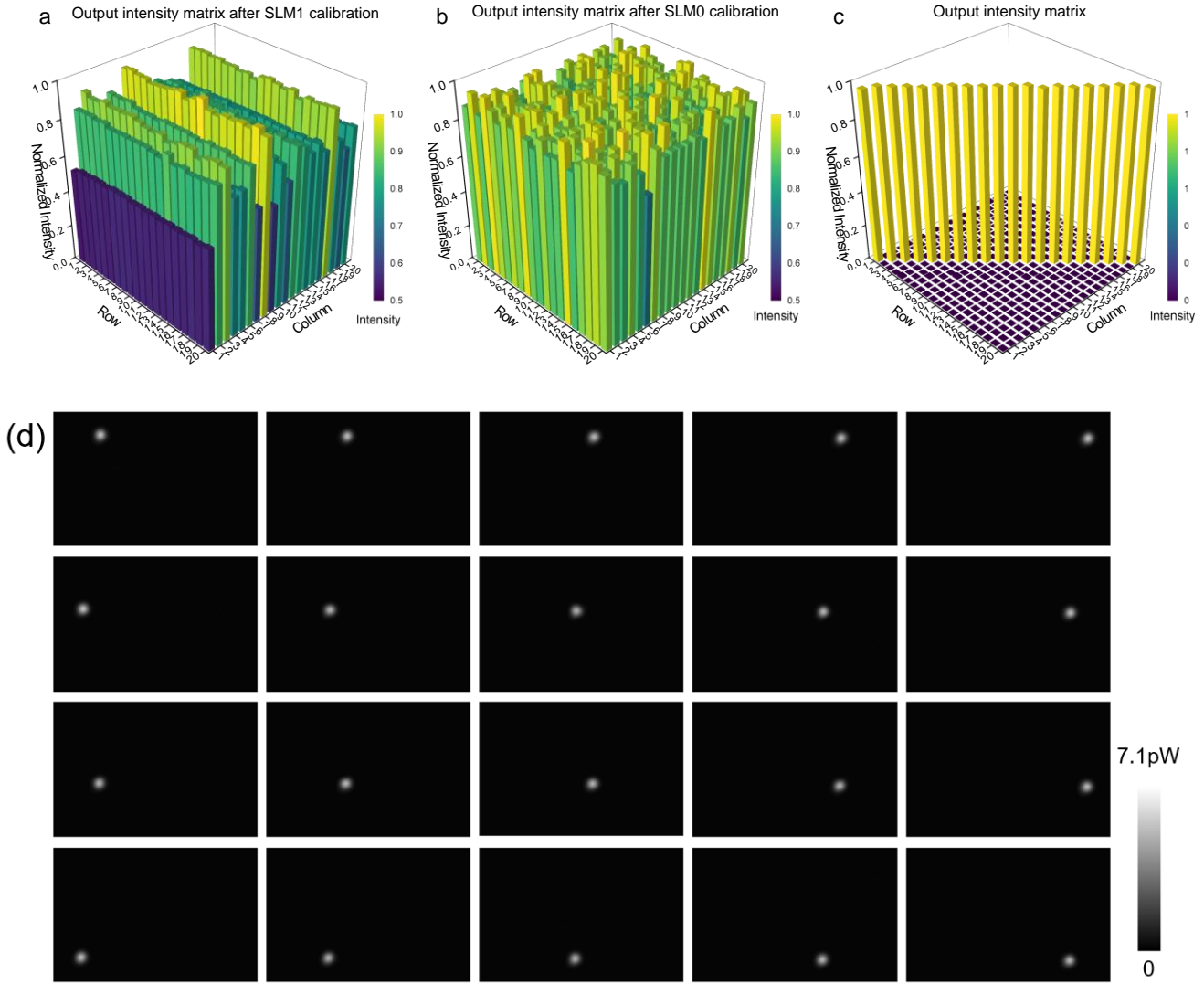

**Supplementary Figure 5**. The calibration of the Discrete Fourier Transformation matrix. (a) The output intensity matrix after the calibration of SLM1. (b) The output intensity matrix after the calibration of SLM0. (c) The output intensity matrix of the calibrated DFT matrix with the input matrix of its inverse matrix. (d) The measured images corresponding to each column in (c) respectively. The color bar denotes the incident power on each pixel.

**Supplementary Note 4: The fidelities of different transformation matrices**

In Supplementary Note 3, the results of the 20-dimensional DFT matrix have verified the validity of our OVMM system. In this section the fidelities of the three matrix transformations of model 1-3 are discussed. Since their inverse matrices usually do not consist of the column vectors with equal norms like DFT matrices, here the output intensity matrices corresponding to the calibrated input identity matrices (like Supplementary Figure 5b) are used to evaluate the transformation fidelities. The results are shown in Supplementary Figure 6. It can be seen that the two matrices are very similar in each model. The fidelities of the intensity matrices are 0.9989, 0.9994, and 0.9969 for model 1-3 respectively. Such results indicate that our OVMM system is also capable of achieving the matrices with highly different elements.

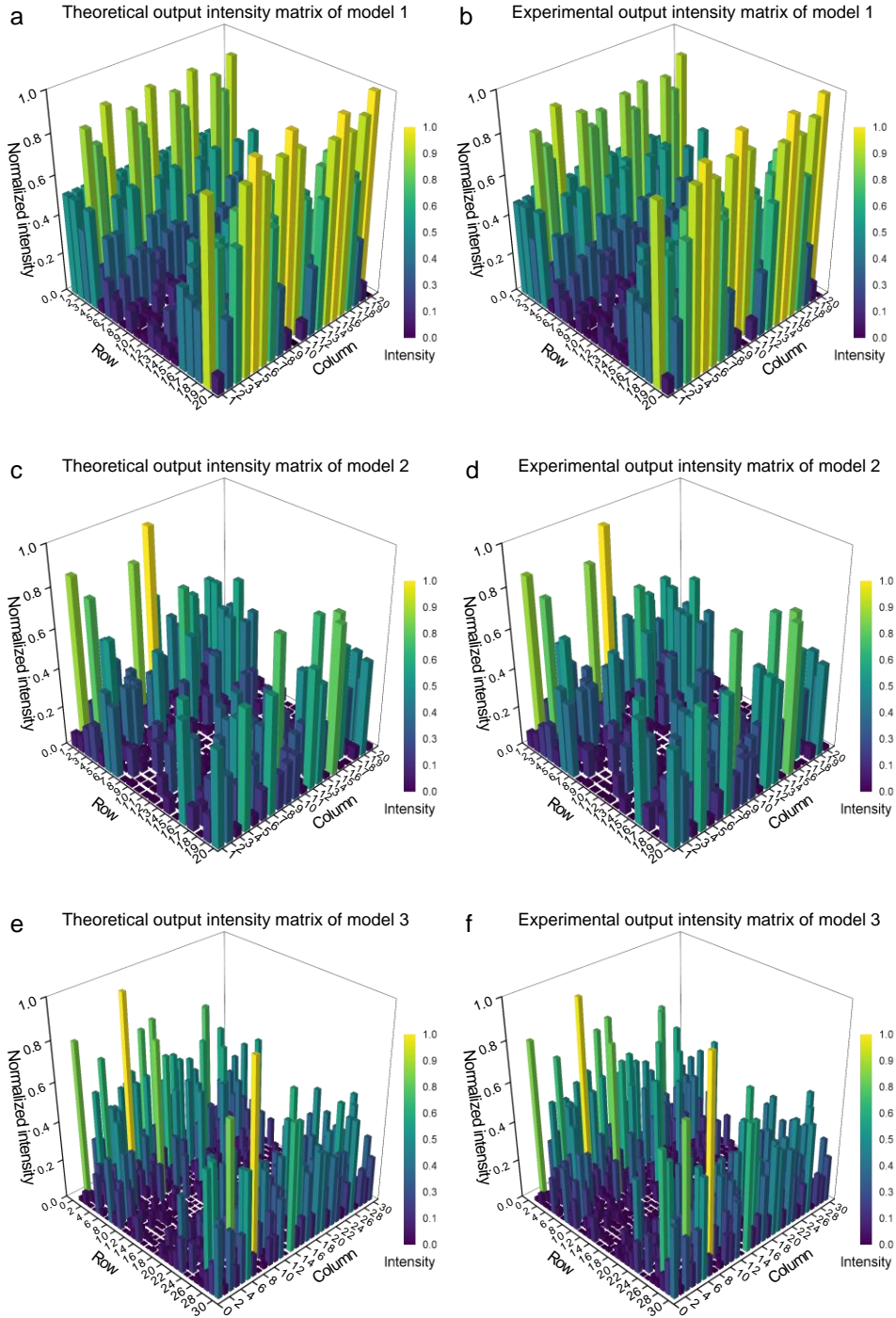

**Supplementary Figure 6**. The comparison of the normalized theoretical and experimental output intensity matrices for model 1-3. (a) (b) The results of model 1. (c) (d) The results of model 2. (e) (f) The results of model 3.

**Supplementary Note 5: The accuracy of the output intensity vectors**

Besides the matrix fidelity, the accuracy of the output vectors also needs to be investigated since the output vectors are directly corresponding to the Hamiltonians. Here, the accuracy of the output vectors is evaluated with two parameters.

According to the main text, there should be a constant normalization coefficient $K = H_{\text{exp}}/H_{\text{theo}}$ (samples with $H_{\text{theo}} = 0$ are excluded) between the theoretical Hamiltonian and the experimental Hamiltonian, which is the first parameter representing the accuracy of the norm of the output intensity vector. The 100 raw experimental Hamiltonian evolution curves obtained from the captured images of model 1-3 are shown in Supplementary Figures 7a-7c respectively. Actually, the target of the PEIDIA is to obtain the spin vector of the ground state, rather than the actual value of the Hamiltonian. Thus, the accepted spin vectors in each iteration corresponding to all curves in Supplementary Figures 7a-7c are extracted to calculate the theoretical Hamiltonians with Eq. (1) in the main text for model 1-3, and the results are shown in Supplementary Figures 7d-7f respectively. From Supplementary Figures 7d-7f it can be seen that most curves converge to the ground state Hamiltonians denoted by the black dashed lines ($H_{\min}$=-26, -62, -117 for model 1-3 respectively) in the end, but their corresponding experimental Hamiltonians are distributed within a small range. To find out the reason, the average value of $K$ of each run of model 1-3 is calculated according to $H_{\text{exp}}$ and $H_{\text{theo}}$ of the actually sampled states rather than the accepted states, and the results are shown in Supplementary Figures 7g-7i respectively. The variation of $K$ in each run is mainly resulted from the detection noise, which will be discussed in detail in Supplementary Note 6. Each experimental Hamiltonian evolution curve is normalized with its own average value of $K$, and the results are shown in Supplementary Figures 7j-7l and for model 1-3 respectively. Such results indicate that our OVMM system possesses relatively high accuracy.

The second parameter is the fidelity $f_{\text{vec}}$ of the output intensity vector, which is defined as

$$f_{\text{vec}} = \frac{\sum_i (I_{\text{exp}})_i (I_{\text{theo}})_i}{\sqrt{\sum_i (I_{\text{exp}})_i^2 \sum_i (I_{\text{theo}})_i^2}}, f_{\text{vec}} \in [0,1], \tag{S30}$$

where $\mathbf{I}_{\text{exp}}$ and $\mathbf{I}_{\text{theo}}$ denote the experimental and theoretical output intensity vectors, respectively. Such fidelity represents the parallelism between the two vectors. In the main text, the vector fidelities of model 1-3 have been discussed with the high average values of 0.99978±0. 00039, 0.99976±0.00044 and 0.99942±0.00131, respectively. Besides, the time stability of the fidelity is also investigated. Supplementary Figure 8 shows the average fidelity of each run of model 1-3 respectively. It can be seen that there is no obvious deterioration of the average fidelities, indicating that our OVMM system is quite stable.

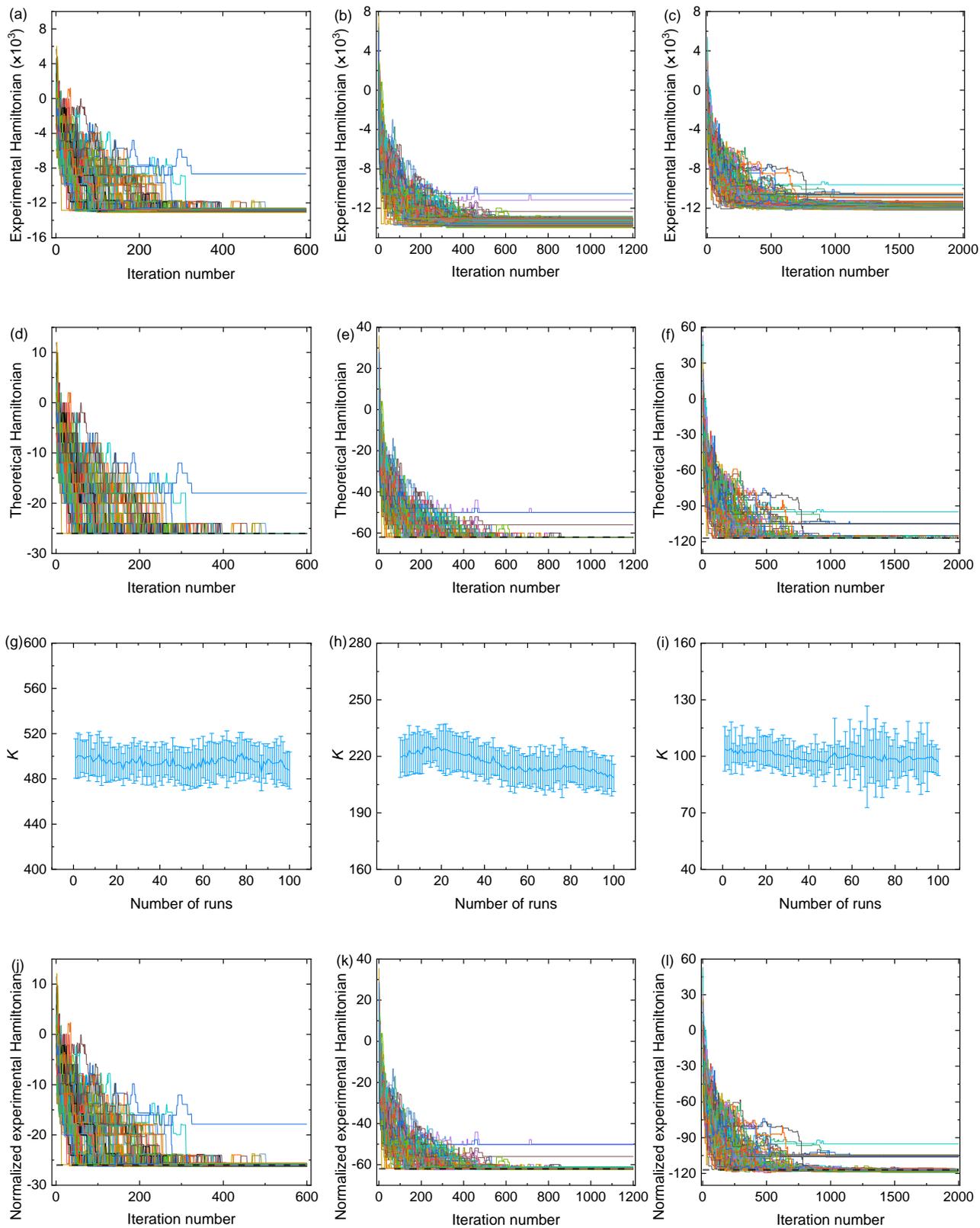

**Supplementary Figure 7**. Results of the ground state search of model 1-3. (a)-(c) The experimental Hamiltonian curves of model 1-3 respectively, which are calculated according to Eq. (S7). (d)-(f) The theoretical Hamiltonian curves corresponding to (a)-(c) respectively. (g)-(i) The average value of $K$ of each run of model 1-3 respectively. The error bar denotes the standard deviation. (j)-(l) The normalized experimental Hamiltonian curves of model 1-3 respectively.

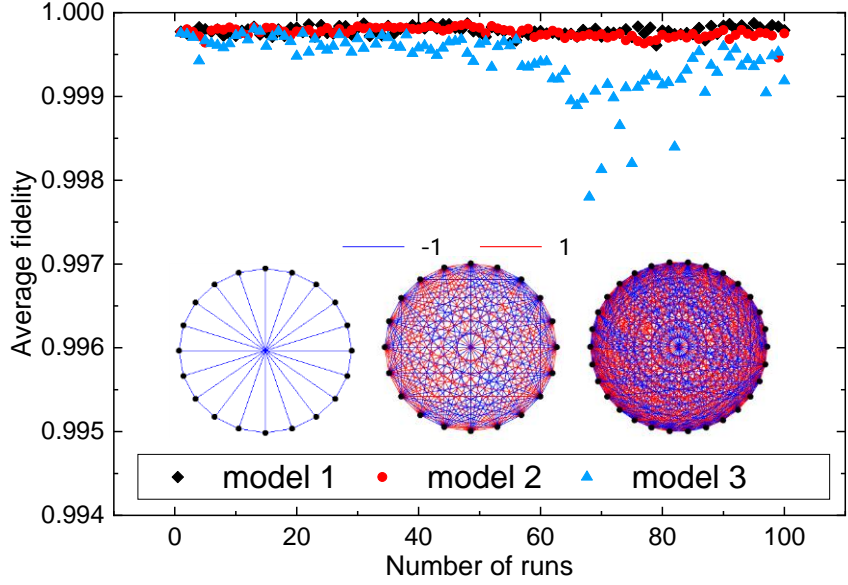

**Supplementary Figure 8**. The average fidelities of 100 runs of model 1-3. The inset shows the corresponding Ising models.

**Supplementary Note 6: The error analysis and the searching accuracy of the experimental Hamiltonians**

Though high transformation fidelities have been achieved in the experiment, the measurement error of the experimental Hamiltonians should be investigated further. According to the experimental results shown in Supplementary Note 4 and 5, the systematic error has been well compensated by the calibration. Thus, we will mainly discuss the random error of measuring Hamiltonians in this section. From Eq. (S7) we have

$$H_{\exp} = \frac{1}{2}\left(\sum_{i,\lambda_i<0} I_i - \sum_{i,\lambda_i>0} I_i\right), \tag{S31}$$

where $H_{\exp}$ is the experimental Hamiltonian, and $I_i$ is the measured intensity of each beam. The uncertainty of $H_{\exp}$ is

$$\delta_H^2 = \sum_{i=1}^N \left(\frac{\partial H_{\exp}}{\partial I_i}\right)^2 \delta_{I_i}^2 = \frac{1}{4}\sum_{i=1}^N \delta_{I_i}^2 \tag{S32}$$

where $N$ is the number of detectors, as well as the optical beams. Here, we assume that the uncertainties introduced by different detectors are independent. The PEIDIA is placed on an optical table with vibration isolators and covered by a glass cabinet, while the temperature of the whole laboratory is kept constant. Besides, the camera has a forced air-cooling system to maintain the operating temperature at 10°C. Thus, it is reasonable to neglect the convection and the drift noise in the experiment, and the uncertainty of each $I_i$ is mainly introduced by the noise of the detector. For the employed Hamamatsu C12741-03 camera, the noise of each pixel $\delta_{I_i}$ mainly consists of 3 independent parts: the readout noise $\delta_r$ (including the thermal noise and the quantization noise $\delta_q$), the dark noise $\delta_d$ and the photon shot noise $\delta_{p_i}$:

$$\delta_{I_i}^2 = \delta_r^2 + \delta_d^2 + \delta_{p_i}^2. \tag{S33}$$

It should be mentioned that the photon shot noise $\delta_{p_i}$ of each pixel is equivalent to the square root of the signal. The noises can be inferred from the parameters in the datasheet of the camera which are listed in Supplementary Table 1:

| Parameter | Symbol | Value | Unit |
|---|---|---|---|
| Full well capacity | $C_{well}$ | $6\times10^5$ | electron |
| Number of bits of the analog-to-digital converter | $B_{ADC}$ | 14 | bit |
| Dark current* (under cooling temperature 10°C) | $i_d$ | 60 | fA |
| Exposure time of a frame | $\Delta t$ | 16.7 | ms |
| Maximum readout noise | $\delta_r$ | 1000 | electron |

**Supplementary Table 1**. The parameters necessary for the noise evaluation. The dark current marked by star is referred from Hamamatsu InGaAs camera G14671-0808W as it is not provided in the datasheet of C12741-03.

In the datasheet, the number of excited electrons per pixel per frame is employed to evaluate the readout noise. Thus, we use the same unit to analysis other noises instead of the common unit of power. Using the parameters in Supplementary Table 1, the noises can be evaluated and listed in Supplementary Table 2.

| Noise type | Symbol | Expression | Value (electron) |
|---|---|---|---|
| Quantization noise | $\delta_q$ | $\dfrac{C_{well}}{2^{B_{ADC}-1}} \cdot \dfrac{1}{\sqrt{12}}$ | 21 |
| Dark noise | $\delta_d$ | $\sqrt{i_d \Delta t / e}$ | 79 |
| Maximum photon shot noise | $\delta_p$ | $\sqrt{C_{well}}$ | 774 |

**Supplementary Table 2**. The expressions and the corresponding values of different types of noise. $e=1.6\times10^{-19}$C is the elementary charge.

It is necessary to investigate the noise level of the ground state, which is our main concern. From the experimentally measured ground state signals of model 1-3 shown in Supplementary Figure 9, it can be found that the signals most concentrates on the last beams respectively.

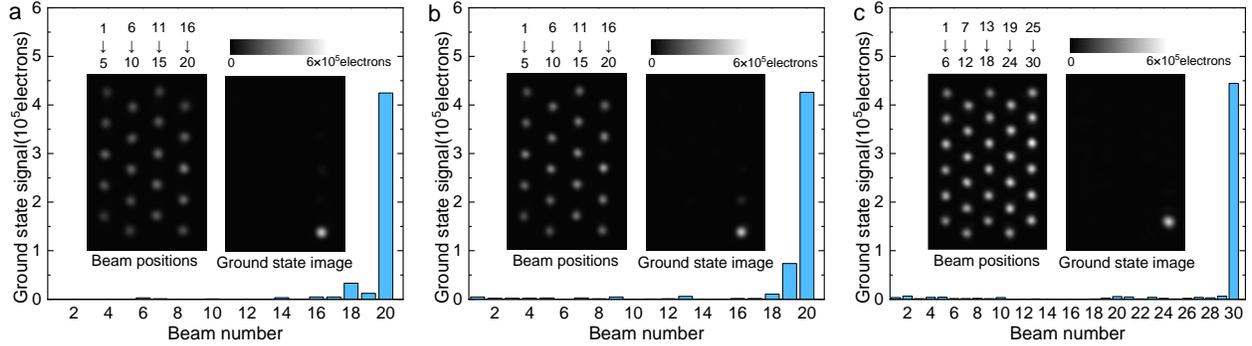

**Supplementary Figure 9**. Experimentally measured ground state signals of model 1-3. (a)-(c) correspond to model 1-3 respectively. The left insets denote the beam positions while the right insets show the ground state images, respectively. The color bar denotes the signal intensity of each pixel in the images.

Thus, we can take a rough approximation of the output signal corresponding to the Ising ground state: the detected signals of beam 1 to $N$-1 are 0 while the signal of beam $N$ is the maximum value $C_{well}$. The corresponding experimental ground state Hamiltonian is $H_0=-3\times10^5$electrons according to Eq. (S31). Thus, the photon shot noises of beam 1 to $N$-1 $\delta_{p_1}, \delta_{p_2}, \ldots, \delta_{p_{N-1}}$ are 0 while that of beam $N$ is $\delta_{p_N} = \sqrt{C_{well}} = 774$electrons. Therefore, the total noises of beam 1 to $N$-1 are $\delta_{I_1} = \delta_{I_2} = \cdots = \delta_{I_{N-1}} = \sqrt{\delta_r^2 + \delta_d^2} \approx 1003$electrons, and that of beam $N$ is $\delta_{I_N} = \sqrt{\delta_r^2 + \delta_d^2 + \delta_{p_N}^2} \approx 1267$electrons. It can be found that the major noise of the whole system is the readout noise of the detector. Then we obtain the uncertainty of the experimental ground state Hamiltonian as

$$\delta_H = \frac{1}{2}\left[(N-1)\delta_{I_1}^2 + \delta_{I_N}^2\right]^{\frac{1}{2}}. \tag{S34}$$

Eq. (S34) indicates that $\delta_\mathrm{H}$ is approximately proportional to $\sqrt{N}$. Moreover, now the signal-to-noise ratio (SNR) of the experimental ground state Hamiltonians can be estimated as $|H_0/\delta_H|$, which are shown in Supplementary Table 3.

| Dimensionality $N$ | $\delta_H$ (electron) | SNR=$20\log|H_0/\delta_H|$ (dB) |
| --- | --- | --- |
| 20 (model 1 and 2) | 2276 | 42.4 |
| 30 (model 3) | 2774 | 40.7 |

**Supplementary Table 3**. The signal-to-noise ratios of the experimental ground state Hamiltonians. $H_0$=-3×10$^5$electrons.

Furthermore, to quantify the influence of the noise during the ground state search, it is reasonable to investigate the "resolution" of the Hamiltonian, which evaluates the searching accuracy of the PEIDIA. An intuitive conclusion is that the PEIDIA cannot distinguish two states if the difference of measured Hamiltonians is less than the noise level. In this discussion, we use the ground state noise level $\delta_H$ to represent the noise level of other spin states as a reasonable approximation. Then we define the resolution of the Hamiltonian by $R = |\delta_H/H_0|$. Besides, we can define the relative Hamiltonian interval $\Delta H_\mathrm{r}$ of two states by the ratio of the difference of the theoretical Hamiltonians $\Delta H_\mathrm{theo}$ to the minimum theoretical Hamiltonian $H_\mathrm{min}$ as $\Delta H_\mathrm{r} = |\Delta H_\mathrm{theo}/H_\mathrm{min}|$. For instance, in model 1-3 the interaction strength $|J_{ij}|$=1. Thus, when only one spin flips, the absolute value of the minimum Hamiltonian variation is 2. Then we obtain the results shown in Supplementary Table 4:

| Parameter | Model 1 | Model 2 | Model 3 |
| --- | --- | --- | --- |
| Minimum theoretical Hamiltonian $H_\mathrm{min}$ | -26 | -62 | -117 |
| Minimum relative Hamiltonian interval $\Delta H_\mathrm{r}$ | 7.69×10$^{-2}$ | 3.22×10$^{-2}$ | 1.71×10$^{-2}$ |
| Resolution of the Hamiltonian $R$ | 7.59×10$^{-3}$ | 7.59×10$^{-3}$ | 9.25×10$^{-3}$ |

**Supplementary Table 4**. The parameters for evaluating the searching accuracy in the experiment of model 1-3.

According to Supplementary Table 4, all $\Delta H_\mathrm{r}$ are larger than $R$ in model 1-3, which indicates that the PEIDIA can well distinguish two states with minimum Hamiltonian interval in model 1-3. Such result is consistent with the achieved high ground state probability in our experiments. The above analysis indicates that the performance of ground state search significantly depends on the Hamiltonian landscape of the given Ising model. The resolution of the Hamiltonian is relative to the dimensionality with the approximation of $R \sim \sqrt{N}$, which indicates the possible deterioration of the searching accuracy in high-dimensional Ising problems. Another interesting conclusion is that although the SNR of $H_\mathrm{exp}$ becomes extremely low when $H_\mathrm{exp} \sim 0$, the PEIDIA can still distinguish different spin states around $H_\mathrm{exp} \sim 0$ since the searching accuracy is only determined by the resolution of the Hamiltonian.

In conclusion, the major noise of the whole system is the readout noise of the detector and the SNR of the ground state Hamiltonian would decrease according to $\sim 1/\sqrt{N}$ with increasing dimensionality $N$, which would be one of the challenges to expand the dimensionality of the PEIDIA.

**Supplementary Note 7: Operation speed and energy consumption**

The time cost per iteration $t_\mathrm{iter}$ in optical domain depends on the propagation time of light $t_\mathrm{p}$, the updating time of the SLM1 $t_\mathrm{u}$ and the detection time of the camera $t_\mathrm{d}$. The distance between the SLM1 and the camera is ~1.5m in the experiment, hence in a single iteration, $t_\mathrm{p}$=5ns. Although the updating of the new SLM1 pattern only needs appending a phase delay mask to the old one, it still requires matrix operations with dimensionality of 1920×1080 (same as the SLM resolution), which takes about $t_\mathrm{u}$=0.15s. Likewise, the transmission and processing of the image (resolution of 230×340) take about $t_\mathrm{d}$=0.17s, while only intensities of $N$ pixels at beam centers are needed. Hence the time cost in the optical domain per iteration is $t_\mathrm{o} = t_\mathrm{p} + t_\mathrm{u} + t_\mathrm{d} \approx 0.32$s. The time cost $t_\mathrm{e}$ of the rest operations of an iteration in the electronic domain is much less than $t_\mathrm{o}$ and is negligible. Thus, the total time cost per iteration is $t_\mathrm{iter} \approx 0.32$s. In each iteration, the OVMM can perform $F = 2N^2 + 2N$ floating-point operations (FLOPs)[11], including $2N^2$ multiplications of $\mathbf{A\sigma}$ and $2N$ multiplications of the element-wise product of the output amplitude vector in the complex field. Take model 3 as an example, for $N$=30, the operation speed of the OVMM is $R = F/t_{\text{iter}}$ =5.81kFLOP/s.

In the experimental demonstration, the power of the input laser is about 2.59μW, 2.07μW and 6.01μW for model 1-3 respectively. The energy consumption mainly depends on the sensitivity of the camera in our current demonstration. With the employed camera, the laser power of $P_l$=6μW is able to achieve high enough signal-to-noise ratio in the experiment of model 3, and the maximum power of the camera is $P_c$=16W. Thus, the energy consumption is $e_{ff} = (P_l + P_c)/R$ =2.75mJ/FLOP.

| Symbol | Description |
| --- | --- |
| $\boldsymbol{\sigma}/\sigma_i$ | Ising spin vector/$i$th Ising spin |
| $H$ | Ising Hamiltonian |
| $N$ | Number of spins in an Ising model |
| $\mathbf{J}/J_{ij}$ | Ising adjacent interaction matrix/element of $\mathbf{J}$ |
| $\mathbf{J}_+$ | The symmetric component of $\mathbf{J}$ |
| $\mathbf{J}_-$ | Anti-symmetric component of $\mathbf{J}$ |
| $\mathbf{Q}$ | Normalized orthogonal eigenvector matrix of $\mathbf{J}$ |
| $\mathbf{D}$ | Diagonal eigenvalue matrix of $\mathbf{J}$ |
| $\lambda_i$ | $i$th eigenvalue of $\mathbf{J}$ |
| $\mathbf{A}$ | $\mathbf{A} = \sqrt{\mathbf{D}}\mathbf{Q}$, the transformation matrix in the OVMM |
| $\mathbf{E}/E_i$ | Output complex amplitude vector /$i$th element of $\mathbf{E}$ |
| $E_0$ | Constant amplitude term of the output field |
| $\varphi_0$ | Constant phase term of the output field |
| $\varphi_i$ | $i$th phase term relative to $\varphi_0$ |
| $\mathbf{I}/I_i$ | Output intensity vector/$i$th output intensity |
| Superscript $\star^{(n)}$ | Certain parameter $\star$ in iteration $n$ |
| $\alpha$ | Cauchy scaling coefficient |
| $T$ | Annealing temperature |
| $\Delta H$ | Hamiltonian variation of two adjacent iteration |
| $f$ | Fidelity of the output intensity vector |
| $\mathbf{I}_{\text{theo}}$ | Ideal output intensity vector |
| $w$ | Radius of the Gaussian beam |
| $w_0$ | Radius of beam waist between SLM1 and SLM2 |
| $\lambda$ | Wavelength of beam |
| $z$ | Axial distance relative to beam waist |
| $z_{\text{SLM}}$ | Half of the distance between SLM1 and SLM2 |
| $w_{\text{SLM}}$ | Minimum beam radius on SLM1 and SLM2 |
| $t_{\text{o}}/t_{\text{e}}$ | Time cost on optical/electronic domain per iteration |
| $t_{\text{p}}$ | Propagation time of light per iteration |
| $t_{\text{u}}$ | Updating time of SLM1 per iteration |
| $t_{\text{d}}$ | Detection time of camera per iteration |
| $F$ | FLOPs of OVMM per iteration |
| $R$ | Operation speed of OVMM |
| $t_{\text{iter}}$ | Total time cost per iteration |
| $P$ | Total energy consumption of the optical system |
| $e_{ff}$ | Energy consumption per FLOP |
| $(T_0)_{\text{exp}}/(T_0)_{\text{simu}}$ | Initial annealing temperature in the experiment/simulation |
| $n_{\text{step}}/n_{\text{temp}}$ | Steps per temperature stage/Stages of temperature |
| $\eta$ | Annealing coefficient |
| $K$ | Normalization coefficient of the experimental Hamiltonian |

**Supplementary Table 5.** Descriptions of the symbols in the main text.